\documentclass[twocolumn,aps,prl]{revtex4}
\usepackage{amsfonts}
\usepackage{amsmath}
\usepackage{amssymb}
\usepackage{graphicx}
\usepackage{color}
\usepackage{ulem}
\usepackage{mathrsfs}
\usepackage[colorlinks, urlcolor=cyan, citecolor=blue, linkcolor=magenta]{hyperref}

\begin{document}

\title{Non-equilibrium phases of Fermi gas inside a cavity \\
with imbalanced pumping}
\author{Xiaotian Nie}
\author{Wei Zheng}
\email{zw8796@ustc.edu.cn}
\affiliation{Hefei National Laboratory for Physical Sciences at the Microscale and
Department of Modern Physics, University of Science and Technology of China,
Hefei 230026, China}
\affiliation{CAS Center for Excellence in Quantum Information and Quantum Physics,
University of Science and Technology of China, Hefei 230026, China}
\affiliation{Hefei National Laboratory, University of Science and Technology of China,
Hefei 230088, China}
\date{\today }

\begin{abstract}
In this work, we investigate the non-equilibrium dynamics of one-dimensional
spinless fermions loaded in a cavity with imbalanced pumping lasers. Our
study is motivated by previous work on a similar setup using bosons, and we
explore the unique properties of fermionic systems in this context. By
considering the imbalance in the pumping, we find that the system exhibits
multiple superradiant steady phases and an unstable phase. Furthermore, by
making use of the hysteresis structure of superradiant phases, we propose a
unidirectional topological pumping. Unlike the usual topological pumping in
which the driving protocol breaks time reversal symmetry, the driving
protocol can be time reversal invariant in our proposal.
\end{abstract}

\maketitle

\section{Introduction}

Ultracold atomic gases coupled to optical cavities provide a versatile
platform for studying quantum many-body physics. On the one hand, cavity photons
mediate long-range interactions between atoms inside a cavity, which can lead
to new phases of atom-cavity hybridized systems. On the other hand, the
leaking of the photons from the cavity provides a dissipation channel that
will drive the system away from equilibrium, exhibiting rich dynamics
and providing a way to detect them. One typical setup involves atoms loaded
into a cavity, which is pumped by a pair of counterpropagating pumping
lasers. Usually, the intensities of the two pumping lasers are balanced,
such that they form a standing wave, and create a static optical lattice for
the atoms. In the past decade, significant advances have been made based on
such balanced pumping setups. For example, superradiance of the cavity field
has been studied and observed with bosonic \cite%
{Dicke@Dicke.1954,Dicke@Hioe.1973,heppSuperradiantPhaseTransition1973,ThyBsSr@Ritsch.2002,ThyBsSr@Vukics.2005,ThyBsSr@Carmichael.2007,ThyBsSr@Domokos.2008,KeldyshDicke@Diehl.2013,ThyBsSr@Ciuti.2014,QFT-OS@Diehl.2016,ThyBsSr@Irish.2017,ThyBsSr@Zilberberg.2018,ThyBsSr@Jia.2020}%
\cite%
{ExpFirstDickeSr@Esslinger.2010,ExpBsSr@Hemmerich.2015,Exp1stNondiss@Esslinger.2021}
and fermionic atoms \cite%
{ThyFmSr@CY.2014,ThyFmSr@Keeling.2014,ThyFmSr@Piazza.2014,ThyFmSr@CY.2015,ThyFmSr@Brennecke.2016,ThyFmSr@Kollath.2016,ThyFmSr@Piazza.2017,ThyFmSr@YW.2018,ThyFmSr@Ritsch.2019}%
\cite{ExpFmSr@WHB.2021} inside cavities respectively. Dissipative time
crystals, which can break discrete or continuous time translation symmetry
\cite%
{ThyBsUns@Keeling.2012,OTC@Piazza.2015,ZW@2016,DickeCTC@Keeling.2018,CTC@Hemmerich.2019,ThyBsUnst@Nunnenkamp.2019,OTC2@Jaksch.2019,OTC@Tuquero.2022,ThyBsUns@Esslinger.2022,Zheng.2023}%
, have also been predicted and observed in such systems. More on-equilibrium
dynamical phases without steady states have been explored in balanced pumped
cavities \cite%
{ExpBsUns@Esslinger.2019,ExpBsUns2@Esslinger.2019,CTC@Hemmerich.2022,ExpBsUns@Esslinger.2022}%
.

Recently, there has been growing interest in exploring the effects of
imbalanced pumping lasers on atom-cavity hybridized systems. The
intensities of the two counterpropagating lasers can be tuned to be unequal,
such that the atoms feel both standing and travelling waves. The asymmetry
in the pumping leads to the emergence of novel phases, including distinct
superradiant phases and self-organized charge pumping \cite%
{Exp1stNondiss@Esslinger.2021,ExpBsUns@Esslinger.2022}. However, though most
of the works are focused on the bosonic atoms inside cavities with
imbalanced pumping, the behavior of fermions in this regime remains largely
unexplored.

In this work, we investigate a one-dimensional cloud of spinless fermions
loaded into an optical cavity and pumped by a pair of transverse laser beams
of unequal intensities. We found new superradiant steady states which did
not appear in the bosonic case, and predicted a self-organized dynamical
phase in such systems. Based on these new superradiant phases, we design a
unidirectional topological pumping. Unlike the usual topological pumping in
which the driving protocol breaks time reversal symmetry, the driving
protocol can be time reversal invariant in our proposal. It is the
self-organization and dissipation that stabilize the quantization of the
pumping. Our work provides insights into the behavior of fermions in cavity
systems, and paves the way for future studies of topological phenomena
inside cavities.

\section{The setup and model}

\begin{figure}[t]
\includegraphics[width=0.45\textwidth]{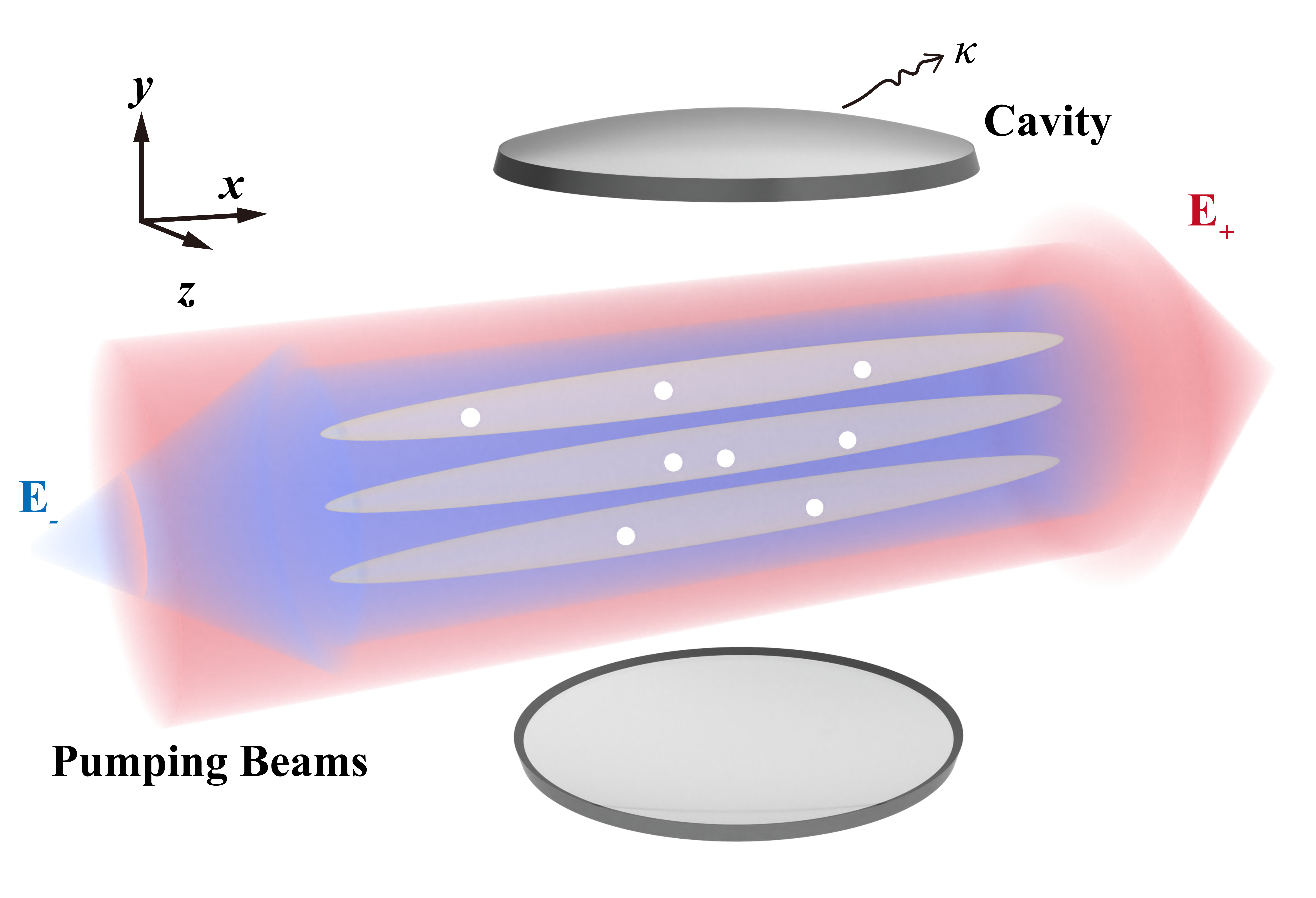}
\caption{A schematic illustration of spinless fermionic atoms trapped in a
cavity coupled with imbalanced transverse pumping beams. The fermion atoms
are restricted in a 1D tube along the pumping beams.}
\label{setup}
\end{figure}

The experiment setup is shown in Fig.\ref{setup}, where fermionic atoms are
loaded into a single-mode optical cavity, which is set along the $y$-axis.
The electrical field of the cavity mode is $\mathbf{\hat{E}}_{\mathrm{c}}(%
\mathbf{r})=\xi (\hat{a}+\hat{a}^{\dagger })\cos (k_{\mathrm{c}}y)\mathbf{e}%
_{z}$, where $\xi $ is the electric field strength of a single photon, and $%
\hat{a}$ ($\hat{a}^{\dagger }$) is the annihilating (creating) operator of
the cavity photons. The atomic cloud is shined by a pair of
counter-propagating pumping lasers along the $x$-axis. The electronic field
of the pumping beams is $\mathbf{E}(\mathbf{r},t)=\mathbf{E}_{+}(\mathbf{r}%
,t)+\mathbf{E}_{-}(\mathbf{r},t)$, with counter-propagating plane waves $%
\mathbf{E}_{\pm }(\mathbf{r},t)=E_{\pm }\cos (\pm {k}_{\mathrm{p}}x-\omega _{%
\mathrm{p}}t)\mathbf{e}_{z}$, where ${k}_{\mathrm{p}}$ is the wave vector of
pumping beam with frequency $\omega _{\mathrm{p}}$.

In such a setup, atoms feel a cavity-dependent potential,%
\begin{equation*}
V(\mathbf{r})=V_{\mathrm{pump}}(\mathbf{r})+V_{\mathrm{cavity}}(\mathbf{r}%
)+V_{\mathrm{inter}}(\mathbf{r}).
\end{equation*}%
Here $V_{\mathrm{pump}}(\mathbf{r})=V_{\mathrm{p}}\cos ^{2}({k}_{\mathrm{p}%
}x)$ is the lattice generated by the pumping lasers, and $V_{\mathrm{p}%
}=u_{s}E_{+}E_{-}/2$ is the corrsponding lattice depth. $V_{\mathrm{cavity}}(%
\mathbf{r})=V_{\mathrm{c}}\cos ^{2}({k}_{\mathrm{c}}y)\hat{a}^{\dagger }\hat{%
a}$ is the lattice generated by the cavity field, and $V_{\mathrm{c}%
}=u_{s}\xi ^{2}$ is the ac Stark shift induced by one cavity photon. The
interference between the pumping beams and cavity field generates the
following lattice,
\begin{eqnarray*}
V_{\mathrm{inter}}(\mathbf{r}) &=&V_{\mathrm{R}}\cos ({k}_{\mathrm{p}}x)\cos
({k}_{\mathrm{c}}y)(\hat{a}+\hat{a}^{\dag })/2 \\
&&+V_{\mathrm{I}}\sin ({k}_{\mathrm{p}}x)\cos ({k}_{\mathrm{c}}y)(\hat{a}-%
\hat{a}^{\dag })/2i,
\end{eqnarray*}%
where $V_{\mathrm{R}}=u_{s}\xi (E_{+}+E_{-})/2$ and $V_{\mathrm{I}}=u_{s}\xi
(E_{+}-E_{-})/2$. It describes the process of scattering a photon by atoms
from pumping lasers into the cavity and vice versa. Here $u_{s}$ is the
scalar polarizability of the atoms. In this work, we only consider blue
atomic detuning, such that $u_{s}>0$. Note that in the case of balanced
pumping, $E_{+}=E_{-}$, thus $V_{\mathrm{I}}=0$, and atoms are only coupled
to the real quadrature of the cavity. When the pumping is imbalanced $%
E_{+}\neq E_{-}$, $V_{\mathrm{I}}\neq 0$, and atoms are coupled to both real
and imaginary quadrature.

In this work, we further consider the motion of atoms to be restricted to
the direction of propagation of the pumping beams. This can be achieved by
adding extra tight trapping potentials along the $x$ direction, such that the
motion of fermions in the $y$ and $z$ directions is frozen, and the system
is effectively one-dimensional. Therefore, the corresponding second
quantized Hamiltonian is given by
\begin{eqnarray}
\hat{H} &=&-\Delta _{\mathrm{c}}\hat{a}^{\dagger }\hat{a}  \notag \\
&&+\int dx~\hat{\psi}^{\dagger }(x)\left[ \frac{-\nabla ^{2}}{2m}+V_{\mathrm{%
p}}\cos ^{2}({k}_{\mathrm{p}}x)\right] \hat{\psi}(x),  \notag \\
&&+\int dx~\hat{\psi}^{\dagger }(x)\left[ V_{\mathrm{R}}\cos ({k}_{\mathrm{p}%
}x)(\hat{a}+\hat{a}^{\dag })/2\right] \hat{\psi}(x)  \notag \\
&&+\int dx~\hat{\psi}^{\dagger }(x)\left[ V_{\mathrm{I}}\sin ({k}_{\mathrm{p}%
}x)(\hat{a}-\hat{a}^{\dag })/2i\right] \hat{\psi}(x)  \label{Ham}
\end{eqnarray}%
where $\hat{\psi}(x)$ is the fermionic field operator of atoms, $\Delta _{%
\mathrm{c}}=\omega _{\mathrm{p}}-\omega _{\mathrm{c}}-NV_{\mathrm{c}}$ is
the effective cavity detuning and $N$ is the total number of femions.

\begin{figure}[t]
\includegraphics[width=0.4\textwidth]{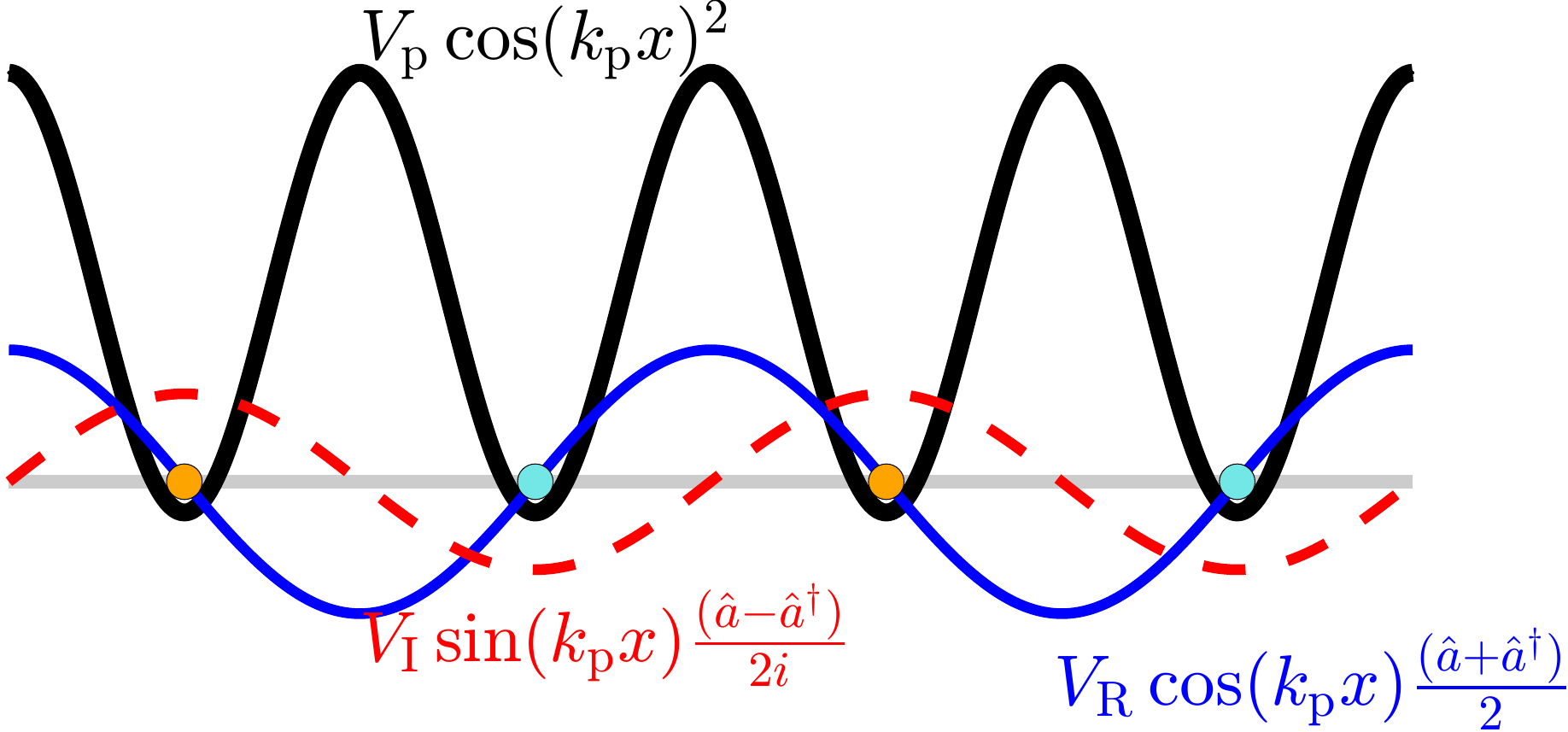}
\caption{Lattice potentials in the continuum. We obtain the tight binding
model by only considering the $s$-band Wannier basis with the nearest
hopping.}
\label{TBM}
\end{figure}

In the strong pumping regime, the lattice generated by the pumping lasers is
so deep, such that the Hamiltonian in the continuous space (\ref{Ham}) can
be simplified into a tight-binding (TB) model. As shown in Fig.\ref{TBM},
the unit cell of the pumping lattice is enlarged due to the double period of
the interference lattice. We denote the two orbits in one unit cell as $%
\mathrm{A}$ and $\mathrm{B}$, and only consider the $s$-band of the pumping
lattice. One obtains the tight-binding Hamiltonian as
\begin{eqnarray}
\hat{H}_{\mathrm{TB}} &=&-\Delta _{\mathrm{c}}\hat{a}^{\dagger }\hat{a}
\notag \\
&&+\sum_{j}J_{0}\left( \hat{c}_{j,\mathrm{B}}^{\dagger }\hat{c}_{j,\mathrm{A}%
}+\hat{c}_{j+1,\mathrm{A}}^{\dagger }\hat{c}_{j,\mathrm{B}}+h.c.\right)
\notag \\
&&+\sum_{j}J_{1}\frac{(\hat{a}+\hat{a}^{\dag })}{2}\left( -\hat{c}_{j,%
\mathrm{B}}^{\dagger }\hat{c}_{j,\mathrm{A}}+\hat{c}_{j+1,\mathrm{A}%
}^{\dagger }\hat{c}_{j,\mathrm{B}}+h.c.\right)  \notag \\
&&+\sum_{j}J_{2}\frac{(\hat{a}-\hat{a}^{\dag })}{2i}\left( \hat{c}_{j,%
\mathrm{A}}^{\dagger }\hat{c}_{j,\mathrm{A}}-\hat{c}_{j,\mathrm{B}}^{\dagger
}\hat{c}_{j,\mathrm{B}}\right) ,
\end{eqnarray}%
where
\begin{eqnarray*}
J_{0} &=&\int_{x}w^{\ast }\left( x-\frac{\lambda _{\mathrm{p}}}{4}\right) %
\left[ \frac{-\nabla ^{2}}{2m}+V_{\mathrm{p}}\cos ^{2}({k}_{\mathrm{p}}x)%
\right] w\left( x+\frac{\lambda _{\mathrm{p}}}{4}\right) , \\
J_{1} &=&V_{\mathrm{R}}\int_{x}w^{\ast }\left( x-\frac{\lambda _{\mathrm{p}}%
}{4}\right) \cos ({k}_{\mathrm{p}}x)w\left( x+\frac{\lambda _{\mathrm{p}}}{4}%
\right) , \\
J_{2} &=&V_{\mathrm{I}}\int_{x}w^{\ast }\left( x-\frac{\lambda _{\mathrm{p}}%
}{4}\right) \sin ({k}_{\mathrm{p}}x)w\left( x-\frac{\lambda _{\mathrm{p}}}{4}%
\right) .
\end{eqnarray*}%
Here $w(x)$ is the $s$-band Wannier wave function in the s-band of the
pumping lattice. Note that this model is a cavity-dependent Rice-Mele model.
The coupling to the real quadrature of the cavity, $(\hat{a}+\hat{a}^{\dag
})/2 $, will tune the hopping ratio between intra- and inter- unit cells,
while coupling to the imaginary quadrature of the cavity, $(\hat{a}-\hat{a}%
^{\dag })/2i$, will change the onsite energy of A/B sublattices. In the
momentum space, the Hamiltonian can be expressed into
\begin{equation*}
\hat{H}_{\mathrm{TB}}=-\Delta _{\mathrm{c}}\hat{a}^{\dagger }\hat{a}%
+\sum\limits_{k}\hat{\Psi}_{k}^{\dag }h(k,\hat{a})\hat{\Psi}_{k},
\end{equation*}%
where $\hat{\Psi}_{k}=(\hat{c}_{k,\mathrm{A}},\hat{c}_{k,\mathrm{B}})^{T}$
and
\begin{equation*}
h(k,\hat{a})=%
\begin{pmatrix}
J_{2}\mathrm{Im}(\hat{a}) & h.c. \\
J_{0}(1+e^{ik})+J_{1}\mathrm{Re}(\hat{a})(-1+e^{ik}) & -J_{2}\mathrm{Im}(%
\hat{a})%
\end{pmatrix}%
.
\end{equation*}

Besides the coherent process governed by the Hamiltonian, the leaking of
photons from the cavity leads to dissipative dynamics. The evolution can be
described by a Lindblad quantum master equation $\partial _{t}\hat{\rho}=-i%
\left[ \hat{H}_{\mathrm{TB}},\hat{\rho}\right] +\kappa \left( 2\hat{a}\hat{%
\rho}\hat{a}^{\dagger }-\left\{ \hat{a}^{\dagger }\hat{a},\hat{\rho}\right\}
\right) $, where $\kappa $ is the photon loss rate.

Apply the mean-field approximation, we obtain the self-consistent
equation-of-motions of the mean cavity field, $\alpha (t)=\left\langle \hat{a%
}(t)\right\rangle $, and fermions as%
\begin{eqnarray}
i\partial _{t}\alpha \left( t\right) &=&(-\Delta _{\mathrm{c}}-i\kappa
)\alpha \left( t\right) +\left\langle \psi \left( t\right) \right\vert \hat{%
\Theta}\left\vert \psi \left( t\right) \right\rangle , \\
i\partial _{t}\left\vert \psi \left( t\right) \right\rangle &=&\hat{H}_{%
\mathrm{MF}}\left[ \alpha (t)\right] \left\vert \psi \left( t\right)
\right\rangle ,
\end{eqnarray}%
where the mean field Hamiltonian of fermions, $\hat{H}_{\mathrm{MF}}\left[
\alpha (t)\right] =\sum\nolimits_{k}\hat{\Psi}_{k}^{\dag }h\left[ k,\alpha
\left( t\right) \right] \hat{\Psi}_{k}$, is dependent on the cavity field $%
\alpha (t)$, and $\hat{\Theta}$ is given by
\begin{equation*}
\hat{\Theta}=\frac{1}{2}\sum\limits_{k}\hat{\Psi}_{k}^{\dag } {\mathbf{\sigma%
}}\cdot%
\begin{pmatrix}
J_{1}(-1+\cos k) \\
J_{1}\sin k \\
iJ_{2}%
\end{pmatrix}%
\hat{\Psi}_{k}.
\end{equation*}%
By solving these equations of motion, one can obtain the dynamics of the
cavity field and fermions.

\begin{figure}[t]
\includegraphics[width=0.4\textwidth]{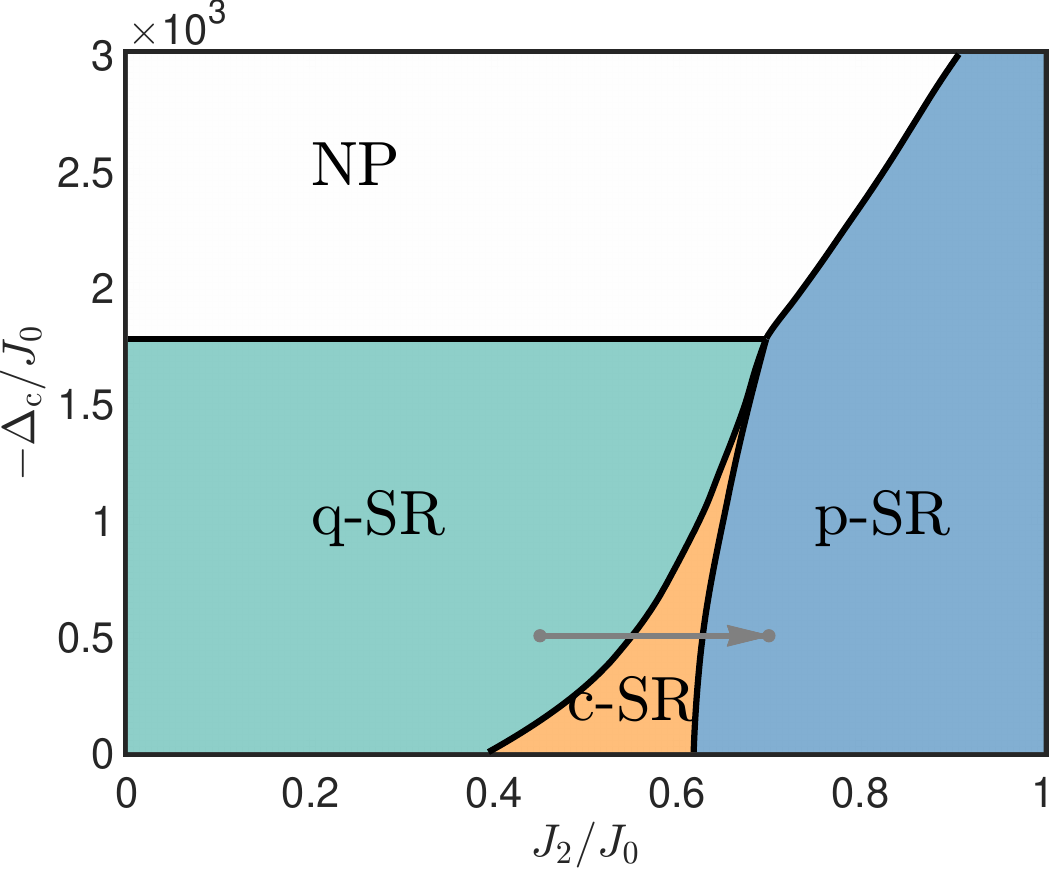}
\caption{Phase diagram without dissipation with $J_{1}/J_{0}=0.5$, filling $%
\protect\nu =0.4$. There are one normal phase(NP) and three superradiant(SR)
phases. Four phases meet at a quadra-critical point. The gray path will be
mentioned in Fig.\protect\ref{TransitionEq}.}
\label{PDEq}
\end{figure}

\section{Non-dissipative case}

In this section, we will first explore this atom-cavity model in the
non-dissipative case $\kappa =0$. This will pave the way to the dissipative
case.

By solving the mean-field equation-of-motions, we obtain the resulting
ground state phase diagram of fixed $J_{1}$ shown in Fig.\ref{PDEq}. The
diagram includes one normal phase (NP) with $\alpha =0$ and three distinct
superradiant (SR) phases $\alpha \neq 0$. When the photon detuning $|\Delta
_{\mathrm{c}}|$ is sufficiently large, the system is in the normal phase. In
the regime of small $|\Delta _{\mathrm{c}}|$ and small $J_{2}$, the system
is in the q-SR phase, which stands for a superradiant phase where the cavity
field $\alpha $ is real. When $J_{2}$ dominates, the system enters the p-SR
phase, where $\alpha $ is purely imaginary. Between them, there is a c-SR
phase where the phase of the cavity field is not fixed, and can be tuned
continuously. The four phases meet at a quadra-critical point.

The symmetry group of the Hamiltonian is product of two $\mathbb{Z}_{2}$
groups, $\{\mathcal{I},\mathcal{T}_{\mathrm{R}}\}\otimes \{\mathcal{I},%
\mathcal{T}_{\mathrm{I}}\}=\{\mathcal{I},\mathcal{T}_{\mathrm{R}},\mathcal{T}%
_{\mathrm{I}},\mathcal{U}\}$ \cite{Ciuti.2012}, where%
\begin{eqnarray*}
\mathcal{T}_{\mathrm{R}} &:&\left\{
\begin{array}{c}
\hat{a}\rightarrow \hat{a}^{\dagger } \\
x\rightarrow -x%
\end{array}%
\right. , \\
\mathcal{T}_{\mathrm{I}} &:&\left\{
\begin{array}{c}
\hat{a}\rightarrow -\hat{a}^{\dagger } \\
x\rightarrow -x+\lambda _{\mathrm{p}}/2%
\end{array}%
\right. , \\
\mathcal{U} &:&\left\{
\begin{array}{c}
\hat{a}\rightarrow -\hat{a} \\
x\rightarrow x+\lambda _{\mathrm{p}}/2%
\end{array}%
\right. .
\end{eqnarray*}%
Note that $\mathcal{T}_{\mathrm{R}}$ represents a real-axis reflection of
cavity field on the complex plane, combining with the spatial reflection of
fermions. $\mathcal{T}_{\mathrm{I}}$ represents an imaginary-axis reflection
with spatial reflection plus spatial translation of fermions. $\mathcal{U=T}%
_{\mathrm{R}}\otimes \mathcal{T}_{\mathrm{I}}$ is a $\pi $-rotation of
cavity field combining a spatial translation of fermions.

\begin{figure}[t]
\includegraphics[width=0.4\textwidth]{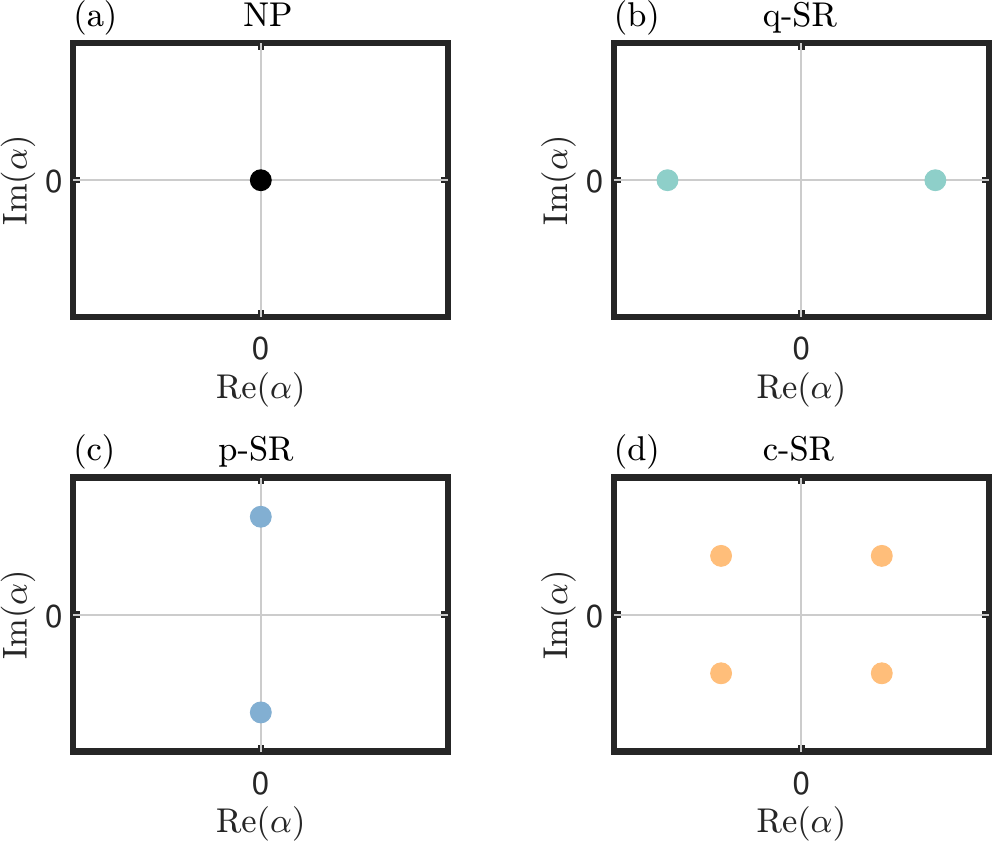}
\caption{Schematic configurations of $\protect\alpha$ of the ground states
on the complex plane in four phases. NP: $\protect\alpha=0$; q-SR: two
opposite real numbers; p-SR: two opposite imaginary numbers; c-SR: four
numbers in four quadrants.}
\label{Eqconfig}
\end{figure}

We show configurations of cavity field in different phases in Fig.\ref%
{Eqconfig} and their symmetries in Table.\ref{tabsymmetry}. All symmetries
are maintained in the NP. In SR phases, including q-SR, p-SR and c-SR
phases, the $\mathcal{U}$ symmetry is broken, but the $\mathcal{T}_{\mathrm{R%
}}$ or $\mathcal{T}_{\mathrm{I}}$ symmetry may survive respectively. The
q-SR phase breaks the $\mathcal{T}_{\mathrm{I}}$ symmetry and keeps the $%
\mathcal{T}_{\mathrm{R}}$ symmetry, thus the phase of cavity is either $0$
or $\pi $. The p-SR phase is invariant under $\mathcal{T}_{\mathrm{I}}$ but
breaks the $\mathcal{T}_{\mathrm{R}}$ symmetry. So the cavity phase is
either $\pi /2$ or $-\pi /2$. The c-SR breaks $\mathcal{T}_{\mathrm{R}}$, $%
\mathcal{T}_{\mathrm{I}}$, and $\mathcal{U}$ symmetries, therefore its
cavity phase can be tuned continuously.

\begin{table}[t]
\caption{Comparison table of symmetries in different phases}
\label{tabsymmetry}\centering
\begin{ruledtabular}
		\begin{tabular}{cccccccc}
		Phase  & arg$(\alpha)$  & $\mathcal{T}_\mathrm{R}$ & $\mathcal{T}_\mathrm{I}$ & $\mathcal{U}$ \\
		\hline
		NP   & \textbackslash & $\checkmark$    & $\checkmark$    & $\checkmark$  \\
		q-SR & $0,\pi$        & $\checkmark$        & $\times$    & $\times$      \\
		p-SR & $\pm\pi/2$     & $\times$    & $\checkmark$        & $\times$      \\
		c-SR & Arbitrary      & $\times$        & $\times$        & $\times$      \\

	\end{tabular}
\end{ruledtabular}
\end{table}

\begin{figure}[b]
\includegraphics[width=0.4\textwidth]{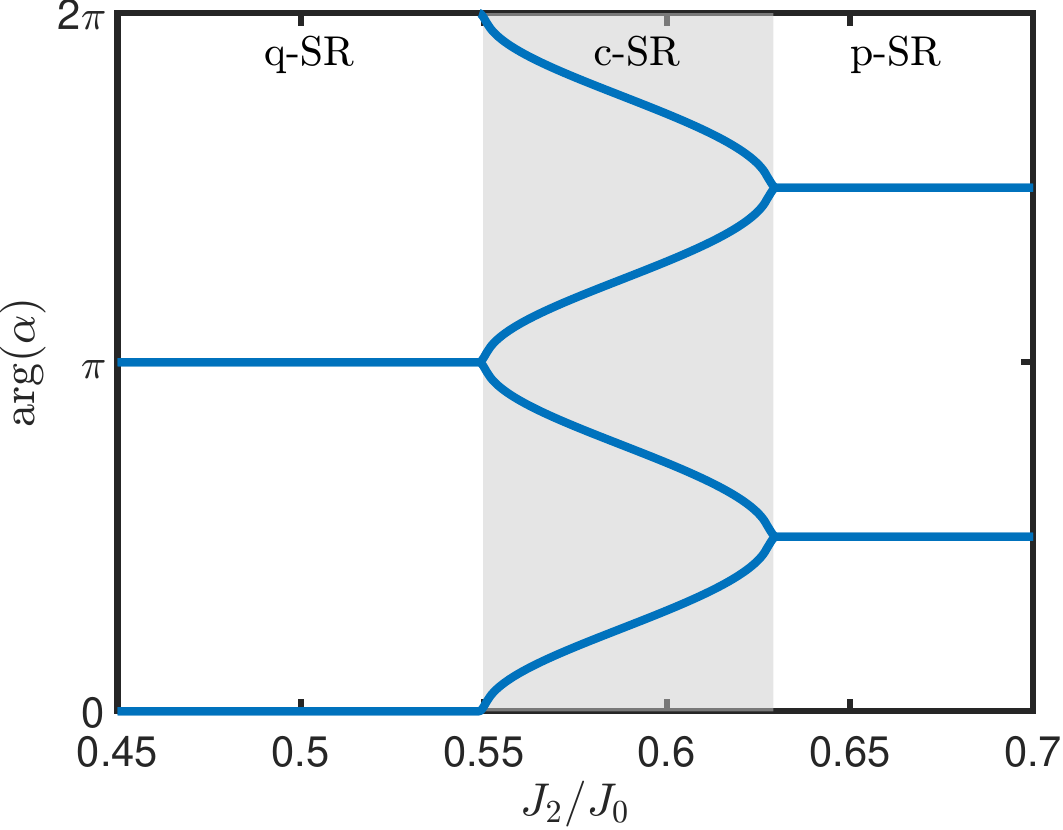}
\caption{The argument of $\protect\alpha $ changes along the gray path in
Fig.\protect\ref{PDEq} acorss the c-SR phases. In the q-SR/p-SR phase, there
are two steady states whose arguments are locked to (0,$\protect\pi $)/$\pm
\frac{\protect\pi }{2}$. In the c-SR phase, the remaining symmetry is
broken, degeneracy is doubled. The transitions are all continuous.}
\label{TransitionEq}
\end{figure}

Next, we investigate the transitions between these phases. We observe that the transitions
are second-order. Starting from NP, by decreasing the $\left\vert \Delta _{%
\mathrm{c}}/J_{0}\right\vert $, the system will enter the q-SR phase. The $%
\mathcal{T}_{\mathrm{I}}$ symmetry is broken spontaneously as two energy
minimums emerge from $\alpha =0$, and then divide oppositely in real axis,
but the $\mathcal{T}_{\mathrm{R}}$ symmetry is preserved. As moving further
into the c-SR phase, the $\mathcal{T}_{\mathrm{R}}$ symmetry is broken by increasing $J_2$, and
each minimum is split into complex conjugate pairs. In the c-SR phase, the
order parameter $\alpha $ changes continuously in four quadrants.
Approaching the transition between the c-SR and p-SR, the upper/lower pair
coalesce into a pure imaginary one respectively, and the $\mathcal{T}_{%
\mathrm{I}}$ symmetry is restored. Finally, the system recovers the $%
\mathcal{T}_{R}$ symmetry by the merging of the imaginary pairs at the
transition to the NP.  Fig.\ref%
{TransitionEq} shows how the phases of cavity field change continuously
along the path depicted in Fig.\ref{PDEq}.

\begin{figure}[t]
\includegraphics[width=0.4\textwidth]{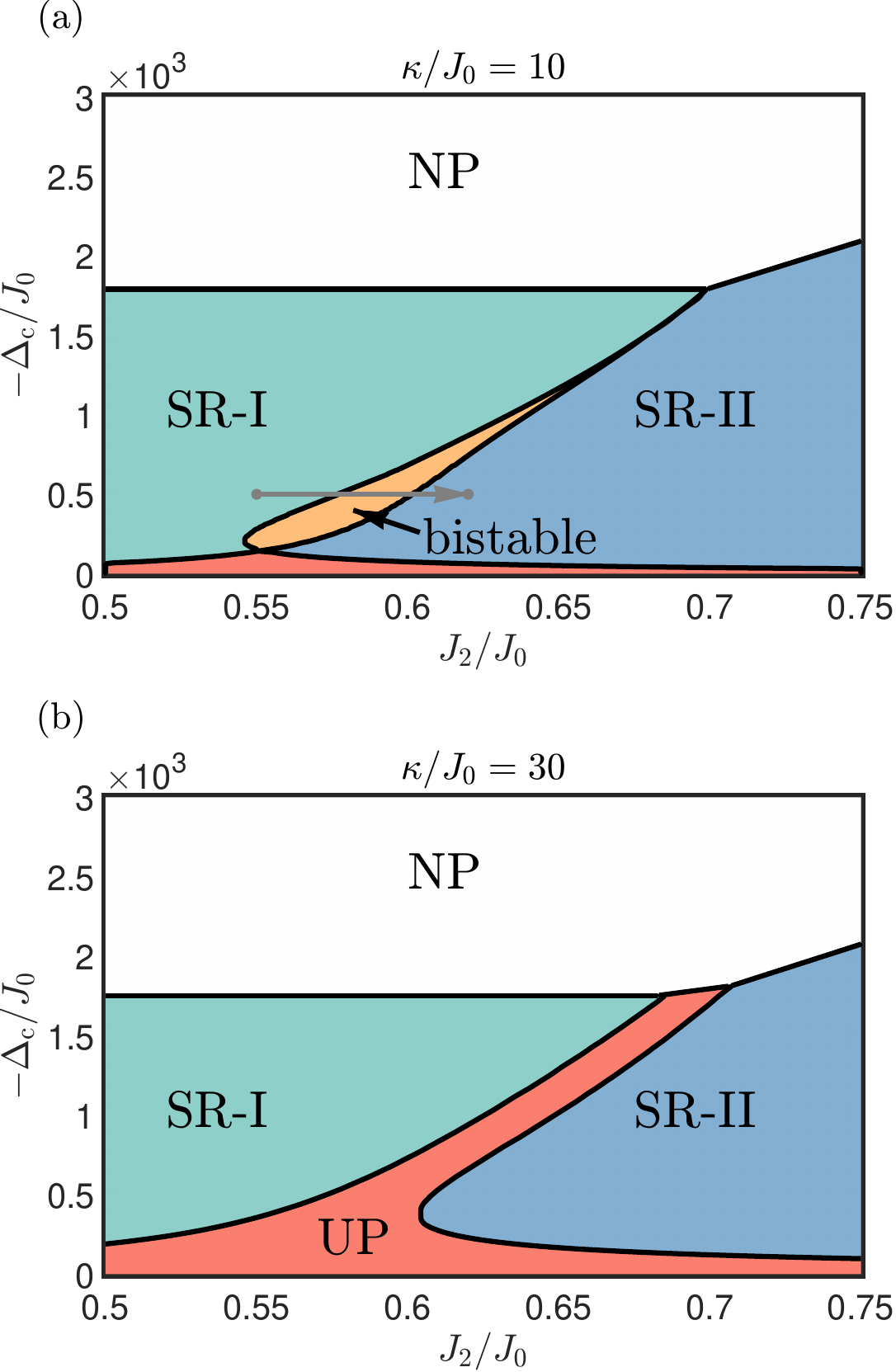}
\caption{The dissipative phase diagram under small and large dissipation
respectively, with parameters $J_{1}/J_{0}=0.5,\protect\nu =0.4$. The new
red region represents the unstable phase. It gradually replaces the bistable
regime as $\protect\kappa $ increases. The gray path will be mentioned in
Fig.\protect\ref{TransitionNonEq}.}
\label{PDNonEq}
\end{figure}

When half filling, the Fermi surface nesting arises due to the interference
lattice coupling the two Fermi momenta \cite%
{ThyFmSr@CY.2014,ThyFmSr@Keeling.2014,ThyFmSr@Piazza.2014}, leading to the
disappearance of the NP. In the q-SR phase, where $\alpha $ is real, the
mean-field Hamiltonian can be simplified to a Su-Schrieffer-Heeger (SSH)
model~\cite{ThyFmSr@Piazza.2017}\cite%
{ThyFmSr@Piazza.2017,ThyFmSr@YW.2018,ThyFmSr@Ritsch.2019}, which has two
topologically distinct phases characterized by the quantized Wannier center~%
\cite{King-Smith.1993,Resta.1994,Niu.2010}.

\section{Dissipative case}

In this section, we will consider the fate of these phases in the presence of
dissipation, $\kappa \neq 0$. In this situation, we numerically solve the
equations of motion, and consider its long-time dynamics to seek the steady
states. The phase diagram in the presence of dissipation is plotted in Fig.%
\ref{PDNonEq}. In the small $\kappa $ regime, there exist five different
regimes: a steady NP and two steady SR phases, which we denote as SR-I and
SR-II. The SR-I phase is reminiscent of the q-SR phase. However, the phase of
the cavity is not locked at $0$ and $\pi $, instead it has a phase shift $%
\phi _{\kappa }=\tan ^{-1}\left( \frac{\Delta _{\mathrm{c}}}{\kappa }\right)
$ relative to the q-SR phase. Similarly, The SR-II phase is reminiscent of
the p-SR phase, but with a phase shift $\phi _{\kappa }$. In the limit $%
\kappa \rightarrow 0$, the SR-I and SR-II phases will continuously crossover
to the q-SR and p-SR phases. From the symmetry point of view, we note that both
the $\mathcal{T}_{\mathrm{R}}$ and $\mathcal{T}_{\mathrm{I}}$ symmetries are
absent in the presence of dissipation. This can be seen from the Lindblad
quantum master equation. The only symmetry survived in the presence of
dissipation is the $\mathcal{U}$ symmetry. Thus, there are no q-SR and p-SR
phases in the dissipative case. Between SR-I and SR-II phases, there is a
bistable regime, in which both SR-I and SR-II phases are the steady state of
the system. Whether the system stays in SR-I or SR-II state, depends on the
initial condition. This bistable regime is reminiscent of the c-SR phase in
the non-dissipative limit. When $|\Delta _{\mathrm{c}}|$ is small, there is
an unstable phase, which does not exist in the non-dissipative case. In this
phase, the system will not reach a steady state. The dissipation will drive
both the cavity field and fermions to evolve incessantly. The unstable phase
has already been observed in bosonic gases coupled with an imbalanced pumped
cavity. In the bosonic case, since there is no bistable regime, the unstable
region emerges directly from the first-order transition between the two
superradiant phases as $\kappa >0$~\cite%
{Exp1stNondiss@Esslinger.2021,ExpBsUns@Esslinger.2022}. Here with fermions
inside an imbalanced pumped cavity, As $\kappa $ increases, we observe that
the unstable region gradually squeezes the bistable regime, eventually
replacing it when $\kappa $ becomes sufficiently large, see Fig.\ref%
{EmergUnstable}.

\begin{figure}[t]
\includegraphics[width=0.4\textwidth]{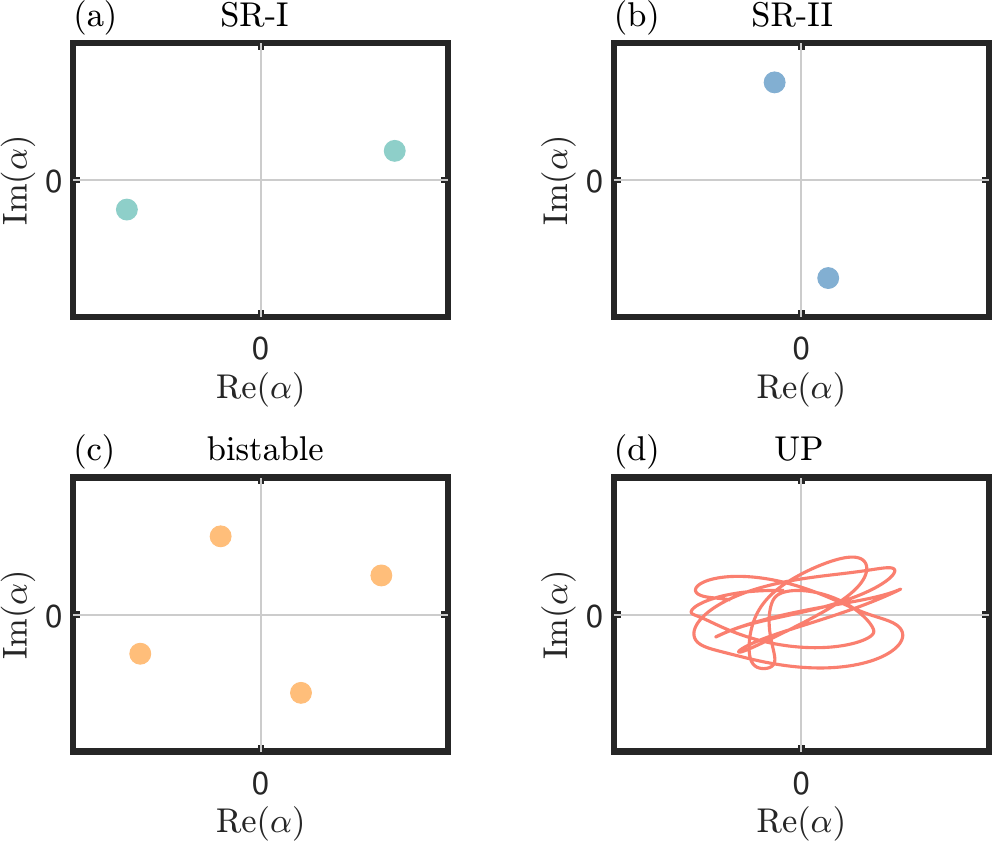}
\caption{Schematic configurations of $\protect\alpha $ of the steady states
on the complex plane in SR phases, and the last figure demonstrates
the trajectory of evolution in the unstable phase.}
\label{nonEqConfig}
\end{figure}

\begin{figure}[t]
\includegraphics[width=0.4\textwidth]{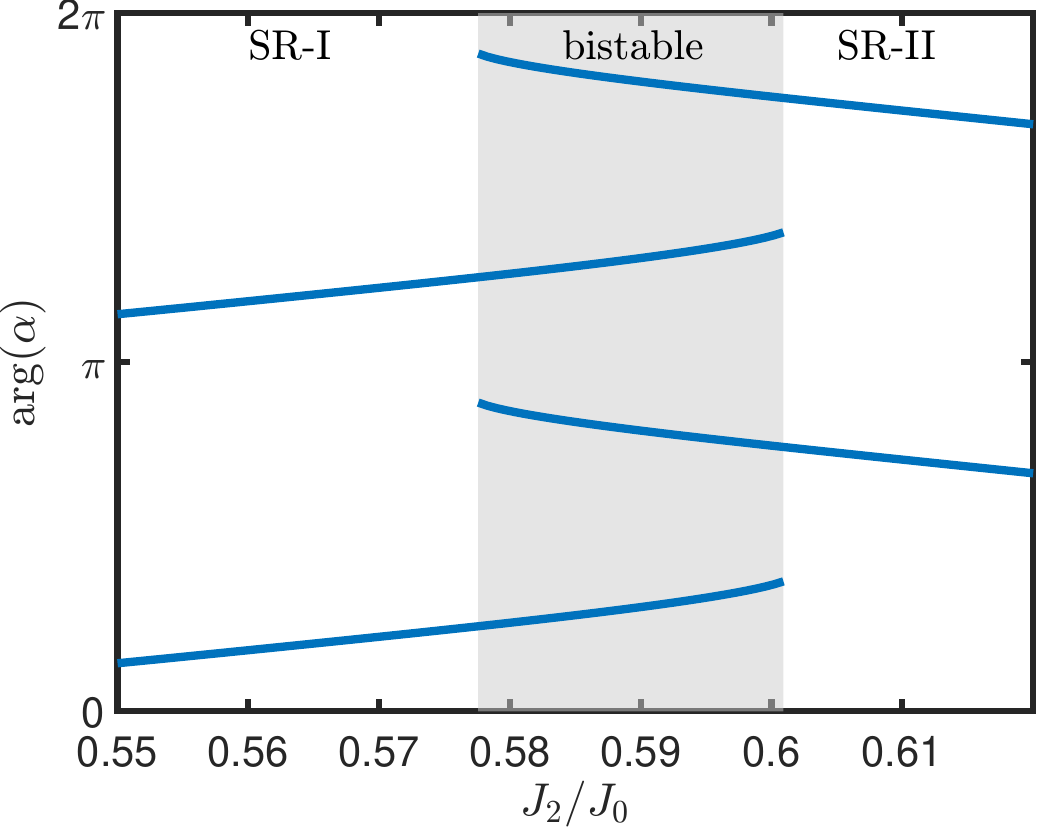}
\caption{The argument of $\protect\alpha$ changes along the gray path in Fig.%
\protect\ref{PDNonEq} acorss the bistable regime. The continuous phase
transitions turn into a hysteresis.}
\label{TransitionNonEq}
\end{figure}

Phase transitions are also strongly influenced by dissipation. In
contrast with the second-order transition in the non-dissipative case, $%
\alpha $ now switches discontinuously when across the bistable regime, and
exhibits a hysteresis structure. When we slowly ramp $J_{2}$ up slowly from
the SR-I, the system will move continuously into the bistable regime, and
will suddenly jump to the SR-II phase at the right boundary. Conversely, if
the system is initially prepared in the SR-II phase, then the jump will take
place on the left boundary when $J_{2}$ is ramped down.

\begin{figure}[b]
\includegraphics[width=0.4\textwidth]{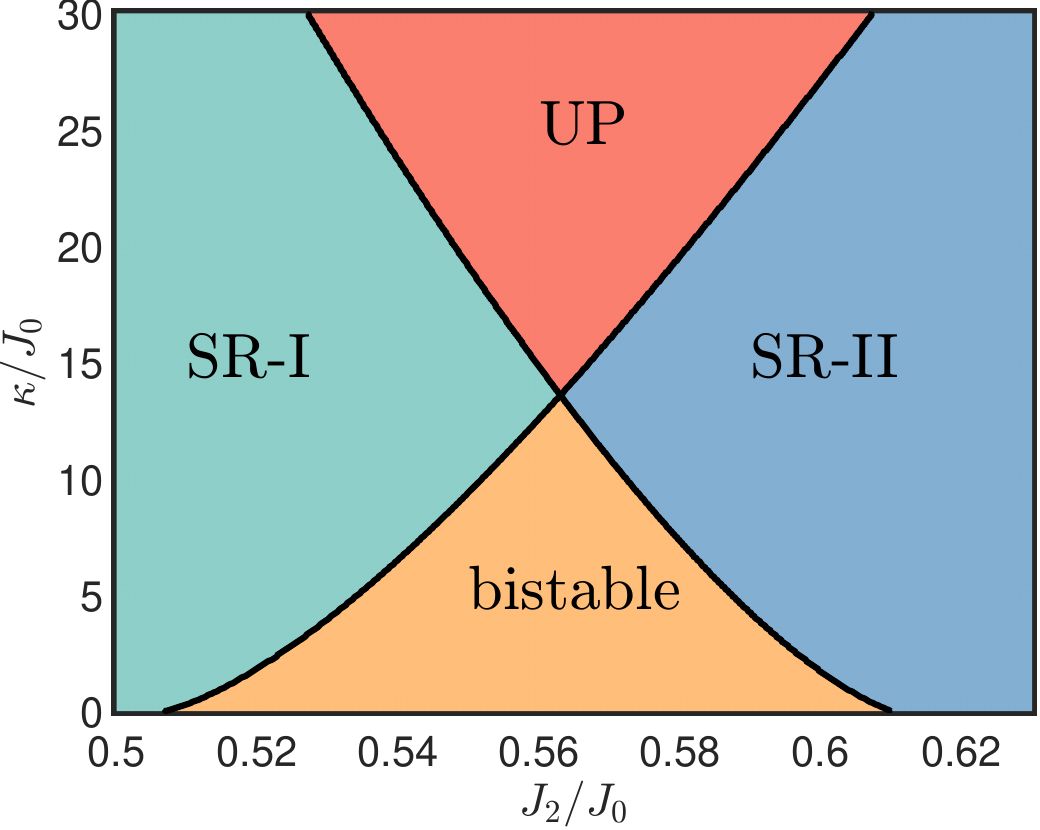}
\caption{A cross-section of the former phase diagrams with fixed $\Delta_%
\mathrm{c}=-300J_0$ but different $\protect\kappa$. The dissipation first
reduces the width of the bistable regime to zero and then brings about an
unstable phase.}
\label{EmergUnstable}
\end{figure}

\begin{figure}[t]
\includegraphics[width=0.4\textwidth]{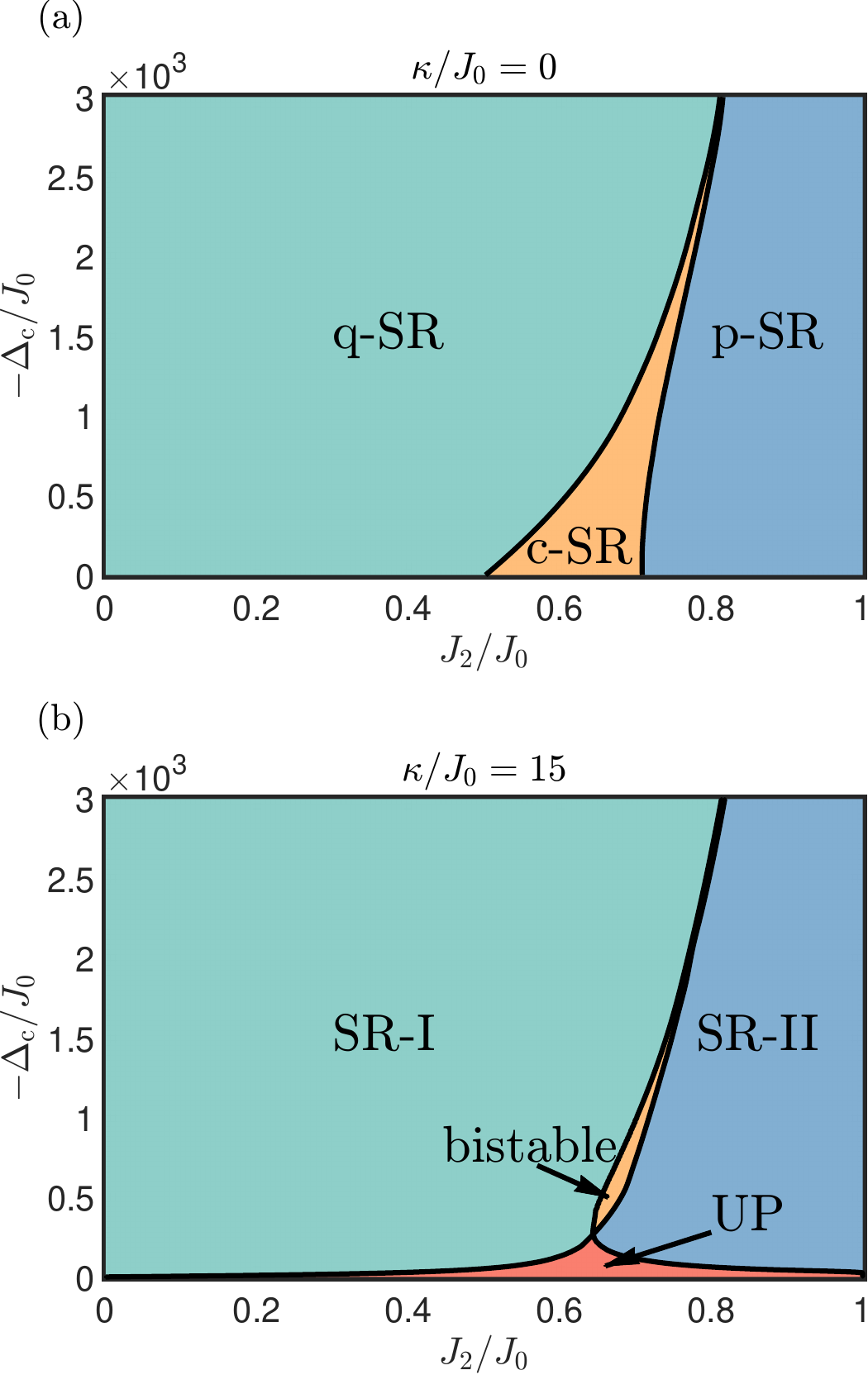}
\caption{Phase diagrams when half filling $\protect\nu=0.5$. The NP is
unfavored due to the Fermi surface nesting. }
\label{PDf0.5}
\end{figure}

\section{Dissipation Induced Unidirectional Topological Pumping}

It has been shown by Thouless that when parameters of a 1D insulator, are
driven adiabatically to complete a cycle, the charge pumped through the bulk
is quantized \cite{Thouless.1983,ThyFmSr@Ritsch.2019,Aidelsburger.2023}. To
pump nonzero quantized charges, the way of the external driving should break
the time reversal symmetry (TRS). For example, the driving protocol could be
chosen as $J_{1}(t)=J_{1}\cos (\Omega t)$ and $J_{2}(t)=J_{2}\sin (\Omega t)$%
. Here, we employ the phase structure of these superradiance states to
realize a unidirectional topological pumping by a TRS-preserved driving
protocol.

\begin{figure}[t]
\includegraphics[width=0.45\textwidth]{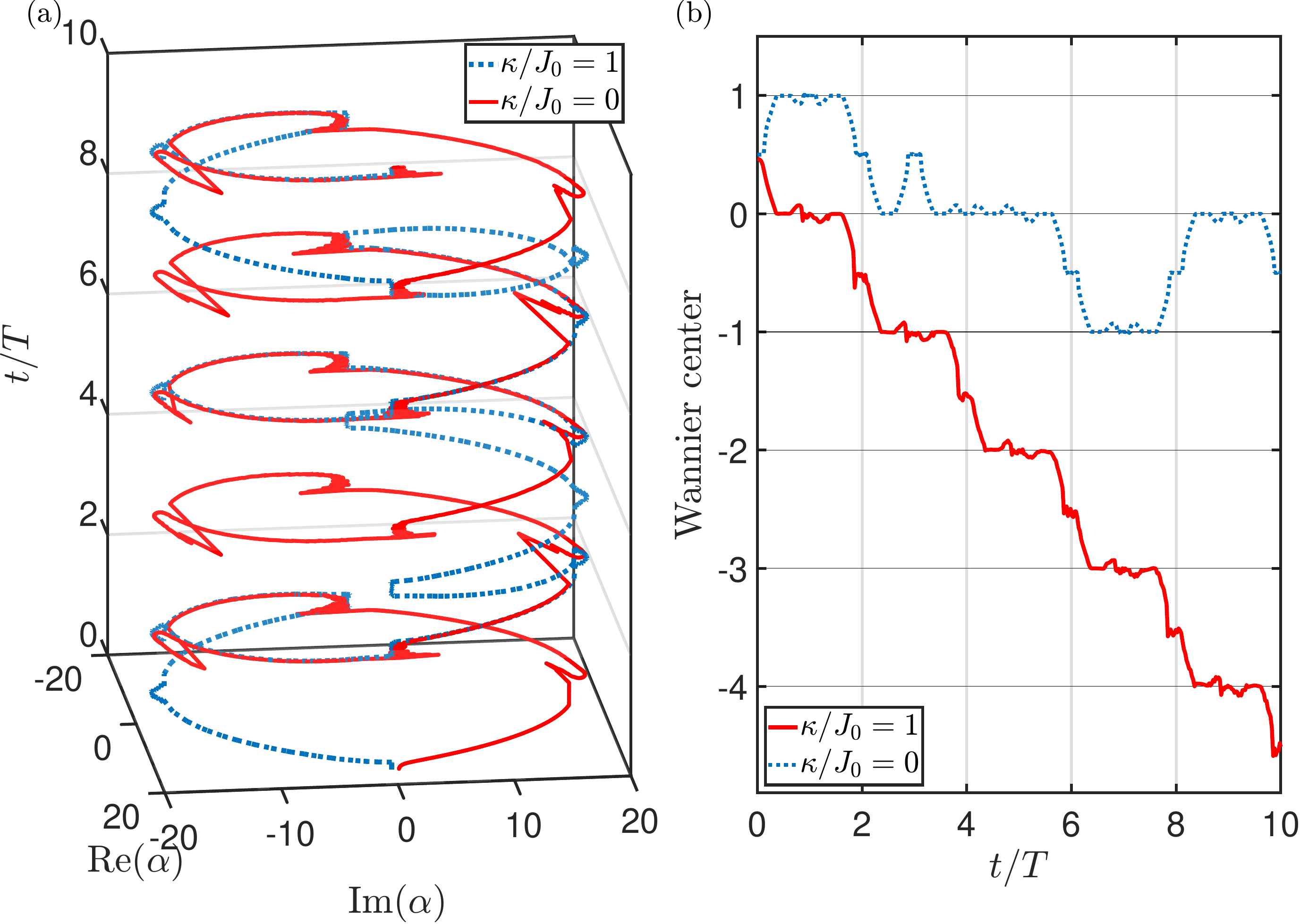} \centering
\caption{The numerical results of time evolution in topological pumping,
where $J_2/J_0 = 0.625-0.125\cos(\frac{2\protect\pi}{T})$ with the period $T
= 2000\protect\pi/J_0 $. (a) and (b) show the time evolution of the absolute
value and argument of $\protect\alpha$, we can see its argument keeps
increasing at a long time scale. And the breaking of the discrete time
translation symmetry can be seen clearly. (c) shows the trajectory of $%
\protect\alpha$ on the complex plane, it forms a circle around the origin,
and has four discontinuous jumps across the axes. The Wannier center flow in
(d) manifests the topological pumping directly. }
\label{pump}
\end{figure}

The numerical results are presented in Fig.\ref{pump}. We perform the
adiabatic driving $J_{2}(t)=J_{2}+\delta J_{2}\cos (\Omega t)$, such that
the instantaneous steady state across the bistable regime with a fixed value
of $\Delta _{\mathrm{c}}$. Our simulations reveal that the evolution of the
cavity field $\alpha (t)$ forms a full circle enclosing the origin on the
complex plane within a doubled period $2T$. That indicates this driven
system exhibits a discrete time crystalline order. In addition, we observe
quantized pumping in the Wannier center trajectory through time evolution
\cite{Takahashi.2016}, see Fig.\ref{pump}.(b).

The physics process under driving is as follows: Starting the driving from
the SR-I phase, the cavity phase is close to $0$, when adiabatic increasing $%
J_{2}$, the system will self-organized follow the driving to enter the
bistable regime smoothly, and stay at one of the stable states. When $J_{2}$
reaches the boundary of the bistable regime, the bistability vanishes, and the
system is forced to jump to the SR-II phase. The excitation energy in this
jump process can be dissipated by the loss of cavity photons, and the system
will catch the SR-II steady state in a short time. In this regime, the
system will again self-organized follow the driving, and the cavity phase is
driven close to $\pi /2$. After half a period, $J_{2}$ will decrease, and the
system will again enter the bistable regime adiabatically, staying at an
alternative stable state. When $J_{2}$ is small, the system will jump to an
SR-I state with cavity phase $\alpha $ close to $\pi $. Repeat the driving
for another period, the cavity field will go back to form a cycle on the
complex plane. In this process, one particle is pumped through the bulk.

\begin{figure}[t]
\includegraphics[width=0.45\textwidth]{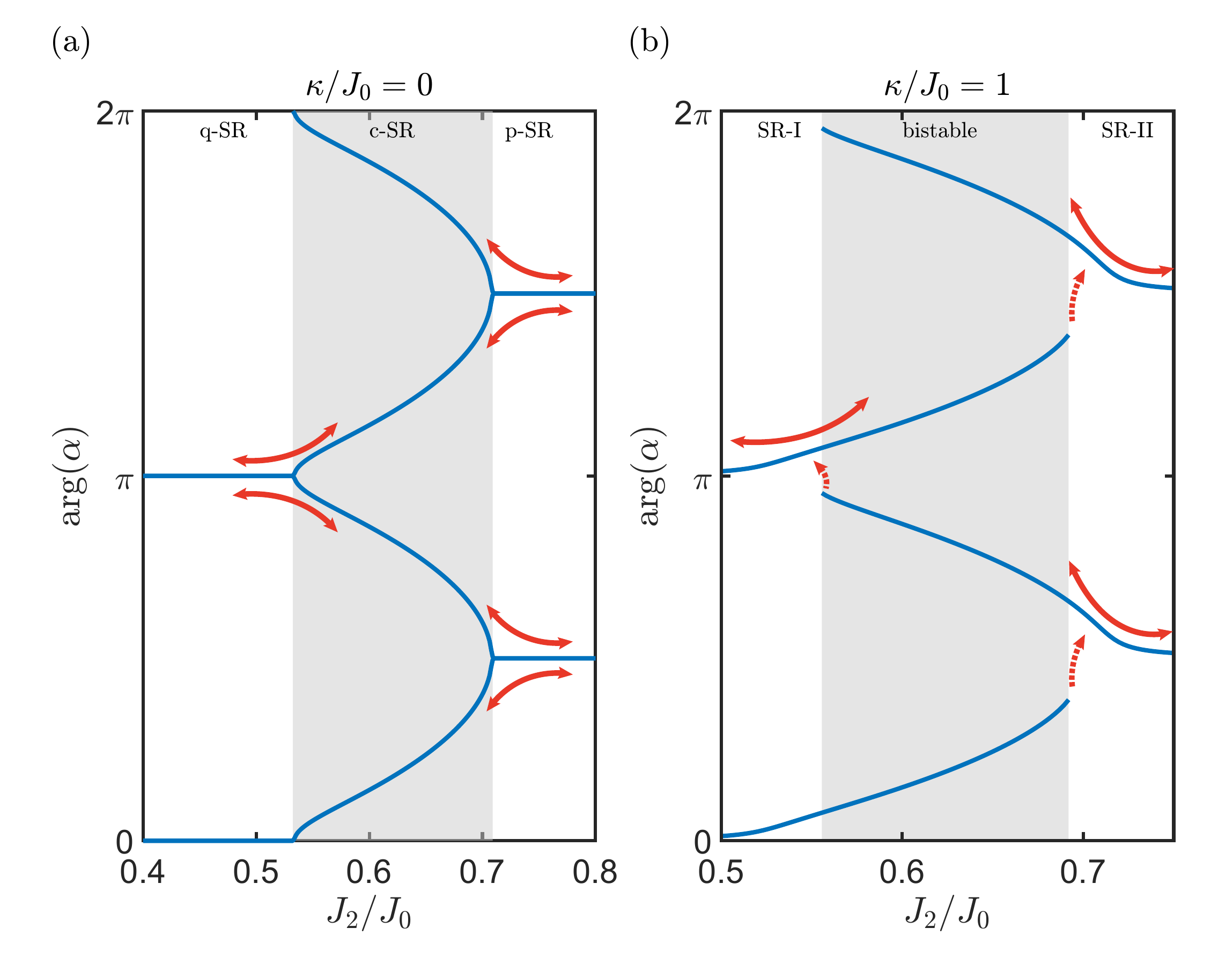} \centering
\caption{The argument of $\protect\alpha$ across the c-SR phase or bistable
regime in the non-dissipative(a) and dissipative(b) case, whose structure
looks like a ladder. The dissipation can turn the bi-directional ladder into
a unidirectional one.}
\label{ladder2}
\end{figure}

We also simulate the same driving protocol without dissipation, $\kappa =0$.
The results are shown in Fig.\ref{pump}. Note that driving the parameter
adiabatically between the q-SR and p-SR phases causes the cavity field $%
\alpha $ to change slowly, resulting in a quantized displacement of the
Wannier center during the evolution. However, the direction of the
displacement is not controllable due to the spontaneous symmetry breaking at
the transitions to the c-SR phase, as shown in Fig.\ref{ladder2}(a). At the
phase boundaries, the phase of $\alpha $ may increase or decrease, leading
to a random back-and-forth displacement of the Wannier center over long
timescales. For a long-time average, the mean displacement is zero.

Compared to the non-dissipative case, we conclude that the dissipation plays
a dual role. Firstly, the presence of dissipation changes the continuous
phase transitions into a bistable hysteresis structure, preventing charge
pumping in both directions. Secondly, it attracts the system towards a
closer steady state while losing stability, which contributes to the
unidirectional motion. This direction of the pumping cannot be reversed by
reversing the driving cycle since the driving protocol preserved the TRS,
but it is the dissipation that breaks the TRS.

\section{Summary}

In summary, we have investigated the behavior of spinless fermions loaded
into an optical cavity and pumped with transverse beams of unequal
intensities. We have observed the emergence of new superradiant phases, and
a novel phenomenon called unidirectional topological pumping.

The study of cavity-coupled ultracold atomic systems continues to be a rich
and exciting field of research, with many intriguing phenomena waiting to be
discovered. We hope that our work will inspire further investigation into
the behavior of these systems and will contribute to the ongoing effort to
understand the behavior of many-body systems far from equilibrium.

\textit{Acknowledgements.} This research is supported by 
the Innovation Program for Quantum Science and Technology (Grant No. 2021ZD0302000).



\begin{thebibliography}{47}%
	\makeatletter
	\providecommand \@ifxundefined [1]{%
	 \@ifx{#1\undefined}
	}%
	\providecommand \@ifnum [1]{%
	 \ifnum #1\expandafter \@firstoftwo
	 \else \expandafter \@secondoftwo
	 \fi
	}%
	\providecommand \@ifx [1]{%
	 \ifx #1\expandafter \@firstoftwo
	 \else \expandafter \@secondoftwo
	 \fi
	}%
	\providecommand \natexlab [1]{#1}%
	\providecommand \enquote  [1]{``#1''}%
	\providecommand \bibnamefont  [1]{#1}%
	\providecommand \bibfnamefont [1]{#1}%
	\providecommand \citenamefont [1]{#1}%
	\providecommand \href@noop [0]{\@secondoftwo}%
	\providecommand \href [0]{\begingroup \@sanitize@url \@href}%
	\providecommand \@href[1]{\@@startlink{#1}\@@href}%
	\providecommand \@@href[1]{\endgroup#1\@@endlink}%
	\providecommand \@sanitize@url [0]{\catcode `\\12\catcode `\$12\catcode
	  `\&12\catcode `\#12\catcode `\^12\catcode `\_12\catcode `\%12\relax}%
	\providecommand \@@startlink[1]{}%
	\providecommand \@@endlink[0]{}%
	\providecommand \url  [0]{\begingroup\@sanitize@url \@url }%
	\providecommand \@url [1]{\endgroup\@href {#1}{\urlprefix }}%
	\providecommand \urlprefix  [0]{URL }%
	\providecommand \Eprint [0]{\href }%
	\providecommand \doibase [0]{http://dx.doi.org/}%
	\providecommand \selectlanguage [0]{\@gobble}%
	\providecommand \bibinfo  [0]{\@secondoftwo}%
	\providecommand \bibfield  [0]{\@secondoftwo}%
	\providecommand \translation [1]{[#1]}%
	\providecommand \BibitemOpen [0]{}%
	\providecommand \bibitemStop [0]{}%
	\providecommand \bibitemNoStop [0]{.\EOS\space}%
	\providecommand \EOS [0]{\spacefactor3000\relax}%
	\providecommand \BibitemShut  [1]{\csname bibitem#1\endcsname}%
	\let\auto@bib@innerbib\@empty
	\bibitem [{\citenamefont {Dicke}(1954)}]{Dicke@Dicke.1954}%
	  \BibitemOpen
	  \bibfield  {author} {\bibinfo {author} {\bibfnamefont {R.~H.}\ \bibnamefont
	  {Dicke}},\ }\href {\doibase 10.1103/PhysRev.93.99} {\bibfield  {journal}
	  {\bibinfo  {journal} {Phys. Rev.}\ }\textbf {\bibinfo {volume} {93}},\
	  \bibinfo {pages} {99} (\bibinfo {year} {1954})}\BibitemShut {NoStop}%
	\bibitem [{\citenamefont {Wang}\ and\ \citenamefont
	  {Hioe}(1973)}]{Dicke@Hioe.1973}%
	  \BibitemOpen
	  \bibfield  {author} {\bibinfo {author} {\bibfnamefont {Y.~K.}\ \bibnamefont
	  {Wang}}\ and\ \bibinfo {author} {\bibfnamefont {F.~T.}\ \bibnamefont
	  {Hioe}},\ }\href {\doibase 10.1103/PhysRevA.7.831} {\bibfield  {journal}
	  {\bibinfo  {journal} {Phys. Rev. A}\ }\textbf {\bibinfo {volume} {7}},\
	  \bibinfo {pages} {831} (\bibinfo {year} {1973})}\BibitemShut {NoStop}%
	\bibitem [{\citenamefont {Hepp}\ and\ \citenamefont
	  {Lieb}(1973)}]{heppSuperradiantPhaseTransition1973}%
	  \BibitemOpen
	  \bibfield  {author} {\bibinfo {author} {\bibfnamefont {K.}~\bibnamefont
	  {Hepp}}\ and\ \bibinfo {author} {\bibfnamefont {E.~H.}\ \bibnamefont
	  {Lieb}},\ }\href {\doibase 10.1016/0003-4916(73)90039-0} {\bibfield
	  {journal} {\bibinfo  {journal} {Annals of Physics}\ }\textbf {\bibinfo
	  {volume} {76}},\ \bibinfo {pages} {360} (\bibinfo {year} {1973})}\BibitemShut
	  {NoStop}%
	\bibitem [{\citenamefont {Domokos}\ and\ \citenamefont
	  {Ritsch}(2002)}]{ThyBsSr@Ritsch.2002}%
	  \BibitemOpen
	  \bibfield  {author} {\bibinfo {author} {\bibfnamefont {P.}~\bibnamefont
	  {Domokos}}\ and\ \bibinfo {author} {\bibfnamefont {H.}~\bibnamefont
	  {Ritsch}},\ }\href {\doibase 10.1103/PhysRevLett.89.253003} {\bibfield
	  {journal} {\bibinfo  {journal} {Phys. Rev. Lett.}\ }\textbf {\bibinfo
	  {volume} {89}},\ \bibinfo {pages} {253003} (\bibinfo {year}
	  {2002})}\BibitemShut {NoStop}%
	\bibitem [{\citenamefont {Asb{\'o}th}\ \emph {et~al.}(2005)\citenamefont
	  {Asb{\'o}th}, \citenamefont {Domokos}, \citenamefont {Ritsch},\ and\
	  \citenamefont {Vukics}}]{ThyBsSr@Vukics.2005}%
	  \BibitemOpen
	  \bibfield  {author} {\bibinfo {author} {\bibfnamefont {J.~K.}\ \bibnamefont
	  {Asb{\'o}th}}, \bibinfo {author} {\bibfnamefont {P.}~\bibnamefont {Domokos}},
	  \bibinfo {author} {\bibfnamefont {H.}~\bibnamefont {Ritsch}}, \ and\ \bibinfo
	  {author} {\bibfnamefont {A.}~\bibnamefont {Vukics}},\ }\href {\doibase
	  10.1103/PhysRevA.72.053417} {\bibfield  {journal} {\bibinfo  {journal} {Phys.
	  Rev. A}\ }\textbf {\bibinfo {volume} {72}},\ \bibinfo {pages} {053417}
	  (\bibinfo {year} {2005})}\BibitemShut {NoStop}%
	\bibitem [{\citenamefont {Dimer}\ \emph {et~al.}(2007)\citenamefont {Dimer},
	  \citenamefont {Estienne}, \citenamefont {Parkins},\ and\ \citenamefont
	  {Carmichael}}]{ThyBsSr@Carmichael.2007}%
	  \BibitemOpen
	  \bibfield  {author} {\bibinfo {author} {\bibfnamefont {F.}~\bibnamefont
	  {Dimer}}, \bibinfo {author} {\bibfnamefont {B.}~\bibnamefont {Estienne}},
	  \bibinfo {author} {\bibfnamefont {A.~S.}\ \bibnamefont {Parkins}}, \ and\
	  \bibinfo {author} {\bibfnamefont {H.~J.}\ \bibnamefont {Carmichael}},\ }\href
	  {\doibase 10.1103/PhysRevA.75.013804} {\bibfield  {journal} {\bibinfo
	  {journal} {Phys. Rev. A}\ }\textbf {\bibinfo {volume} {75}},\ \bibinfo
	  {pages} {013804} (\bibinfo {year} {2007})}\BibitemShut {NoStop}%
	\bibitem [{\citenamefont {Nagy}\ \emph {et~al.}(2008)\citenamefont {Nagy},
	  \citenamefont {Szirmai},\ and\ \citenamefont
	  {Domokos}}]{ThyBsSr@Domokos.2008}%
	  \BibitemOpen
	  \bibfield  {author} {\bibinfo {author} {\bibfnamefont {D.}~\bibnamefont
	  {Nagy}}, \bibinfo {author} {\bibfnamefont {G.}~\bibnamefont {Szirmai}}, \
	  and\ \bibinfo {author} {\bibfnamefont {P.}~\bibnamefont {Domokos}},\ }\href
	  {\doibase 10.1140/epjd/e2008-00074-6} {\bibfield  {journal} {\bibinfo
	  {journal} {Eur. Phys. J. D}\ }\textbf {\bibinfo {volume} {48}},\ \bibinfo
	  {pages} {127} (\bibinfo {year} {2008})}\BibitemShut {NoStop}%
	\bibitem [{\citenamefont {Torre}\ \emph {et~al.}(2013)\citenamefont {Torre},
	  \citenamefont {Diehl}, \citenamefont {Lukin}, \citenamefont {Sachdev},\ and\
	  \citenamefont {Strack}}]{KeldyshDicke@Diehl.2013}%
	  \BibitemOpen
	  \bibfield  {author} {\bibinfo {author} {\bibfnamefont {E.~G.~D.}\
	  \bibnamefont {Torre}}, \bibinfo {author} {\bibfnamefont {S.}~\bibnamefont
	  {Diehl}}, \bibinfo {author} {\bibfnamefont {M.~D.}\ \bibnamefont {Lukin}},
	  \bibinfo {author} {\bibfnamefont {S.}~\bibnamefont {Sachdev}}, \ and\
	  \bibinfo {author} {\bibfnamefont {P.}~\bibnamefont {Strack}},\ }\href
	  {\doibase 10.1103/PhysRevA.87.023831} {\bibfield  {journal} {\bibinfo
	  {journal} {Phys. Rev. A}\ }\textbf {\bibinfo {volume} {87}},\ \bibinfo
	  {pages} {023831} (\bibinfo {year} {2013})}\BibitemShut {NoStop}%
	\bibitem [{\citenamefont {Baksic}\ and\ \citenamefont
	  {Ciuti}(2014)}]{ThyBsSr@Ciuti.2014}%
	  \BibitemOpen
	  \bibfield  {author} {\bibinfo {author} {\bibfnamefont {A.}~\bibnamefont
	  {Baksic}}\ and\ \bibinfo {author} {\bibfnamefont {C.}~\bibnamefont {Ciuti}},\
	  }\href {\doibase 10.1103/PhysRevLett.112.173601} {\bibfield  {journal}
	  {\bibinfo  {journal} {Phys. Rev. Lett.}\ }\textbf {\bibinfo {volume} {112}},\
	  \bibinfo {pages} {173601} (\bibinfo {year} {2014})}\BibitemShut {NoStop}%
	\bibitem [{\citenamefont {Sieberer}\ \emph {et~al.}(2016)\citenamefont
	  {Sieberer}, \citenamefont {Buchhold},\ and\ \citenamefont
	  {Diehl}}]{QFT-OS@Diehl.2016}%
	  \BibitemOpen
	  \bibfield  {author} {\bibinfo {author} {\bibfnamefont {L.~M.}\ \bibnamefont
	  {Sieberer}}, \bibinfo {author} {\bibfnamefont {M.}~\bibnamefont {Buchhold}},
	  \ and\ \bibinfo {author} {\bibfnamefont {S.}~\bibnamefont {Diehl}},\ }\href
	  {\doibase 10.1088/0034-4885/79/9/096001} {\bibfield  {journal} {\bibinfo
	  {journal} {Rep. Prog. Phys.}\ }\textbf {\bibinfo {volume} {79}},\ \bibinfo
	  {pages} {096001} (\bibinfo {year} {2016})}\BibitemShut {NoStop}%
	\bibitem [{\citenamefont {Larson}\ and\ \citenamefont
	  {Irish}(2017)}]{ThyBsSr@Irish.2017}%
	  \BibitemOpen
	  \bibfield  {author} {\bibinfo {author} {\bibfnamefont {J.}~\bibnamefont
	  {Larson}}\ and\ \bibinfo {author} {\bibfnamefont {E.~K.}\ \bibnamefont
	  {Irish}},\ }\href {\doibase 10.1088/1751-8121/aa65dc} {\bibfield  {journal}
	  {\bibinfo  {journal} {J. Phys. A: Math. Theor.}\ }\textbf {\bibinfo {volume}
	  {50}},\ \bibinfo {pages} {174002} (\bibinfo {year} {2017})}\BibitemShut
	  {NoStop}%
	\bibitem [{\citenamefont {Soriente}\ \emph {et~al.}(2018)\citenamefont
	  {Soriente}, \citenamefont {Donner}, \citenamefont {Chitra},\ and\
	  \citenamefont {Zilberberg}}]{ThyBsSr@Zilberberg.2018}%
	  \BibitemOpen
	  \bibfield  {author} {\bibinfo {author} {\bibfnamefont {M.}~\bibnamefont
	  {Soriente}}, \bibinfo {author} {\bibfnamefont {T.}~\bibnamefont {Donner}},
	  \bibinfo {author} {\bibfnamefont {R.}~\bibnamefont {Chitra}}, \ and\ \bibinfo
	  {author} {\bibfnamefont {O.}~\bibnamefont {Zilberberg}},\ }\href {\doibase
	  10.1103/PhysRevLett.120.183603} {\bibfield  {journal} {\bibinfo  {journal}
	  {Phys. Rev. Lett.}\ }\textbf {\bibinfo {volume} {120}},\ \bibinfo {pages}
	  {183603} (\bibinfo {year} {2018})}\BibitemShut {NoStop}%
	\bibitem [{\citenamefont {Fan}\ \emph {et~al.}(2020)\citenamefont {Fan},
	  \citenamefont {Chen},\ and\ \citenamefont {Jia}}]{ThyBsSr@Jia.2020}%
	  \BibitemOpen
	  \bibfield  {author} {\bibinfo {author} {\bibfnamefont {J.}~\bibnamefont
	  {Fan}}, \bibinfo {author} {\bibfnamefont {G.}~\bibnamefont {Chen}}, \ and\
	  \bibinfo {author} {\bibfnamefont {S.}~\bibnamefont {Jia}},\ }\href {\doibase
	  10.1103/PhysRevA.101.063627} {\bibfield  {journal} {\bibinfo  {journal}
	  {Phys. Rev. A}\ }\textbf {\bibinfo {volume} {101}},\ \bibinfo {pages}
	  {063627} (\bibinfo {year} {2020})}\BibitemShut {NoStop}%
	\bibitem [{\citenamefont {Baumann}\ \emph {et~al.}(2010)\citenamefont
	  {Baumann}, \citenamefont {Guerlin}, \citenamefont {Brennecke},\ and\
	  \citenamefont {Esslinger}}]{ExpFirstDickeSr@Esslinger.2010}%
	  \BibitemOpen
	  \bibfield  {author} {\bibinfo {author} {\bibfnamefont {K.}~\bibnamefont
	  {Baumann}}, \bibinfo {author} {\bibfnamefont {C.}~\bibnamefont {Guerlin}},
	  \bibinfo {author} {\bibfnamefont {F.}~\bibnamefont {Brennecke}}, \ and\
	  \bibinfo {author} {\bibfnamefont {T.}~\bibnamefont {Esslinger}},\ }\href
	  {\doibase 10.1038/nature09009} {\bibfield  {journal} {\bibinfo  {journal}
	  {Nature}\ }\textbf {\bibinfo {volume} {464}},\ \bibinfo {pages} {1301}
	  (\bibinfo {year} {2010})}\BibitemShut {NoStop}%
	\bibitem [{\citenamefont {Klinder}\ \emph {et~al.}(2015)\citenamefont
	  {Klinder}, \citenamefont {Ke{\ss}ler}, \citenamefont {Wolke}, \citenamefont
	  {Mathey},\ and\ \citenamefont {Hemmerich}}]{ExpBsSr@Hemmerich.2015}%
	  \BibitemOpen
	  \bibfield  {author} {\bibinfo {author} {\bibfnamefont {J.}~\bibnamefont
	  {Klinder}}, \bibinfo {author} {\bibfnamefont {H.}~\bibnamefont {Ke{\ss}ler}},
	  \bibinfo {author} {\bibfnamefont {M.}~\bibnamefont {Wolke}}, \bibinfo
	  {author} {\bibfnamefont {L.}~\bibnamefont {Mathey}}, \ and\ \bibinfo {author}
	  {\bibfnamefont {A.}~\bibnamefont {Hemmerich}},\ }\href {\doibase
	  10.1073/pnas.1417132112} {\bibfield  {journal} {\bibinfo  {journal} {Proc.
	  Natl. Acad. Sci. U.S.A.}\ }\textbf {\bibinfo {volume} {112}},\ \bibinfo
	  {pages} {3290} (\bibinfo {year} {2015})}\BibitemShut {NoStop}%
	\bibitem [{\citenamefont {Li}\ \emph {et~al.}(2021)\citenamefont {Li},
	  \citenamefont {Dreon}, \citenamefont {Zupancic}, \citenamefont
	  {Baumg{\"a}rtner}, \citenamefont {Morales}, \citenamefont {Zheng},
	  \citenamefont {Cooper}, \citenamefont {Donner},\ and\ \citenamefont
	  {Esslinger}}]{Exp1stNondiss@Esslinger.2021}%
	  \BibitemOpen
	  \bibfield  {author} {\bibinfo {author} {\bibfnamefont {X.}~\bibnamefont
	  {Li}}, \bibinfo {author} {\bibfnamefont {D.}~\bibnamefont {Dreon}}, \bibinfo
	  {author} {\bibfnamefont {P.}~\bibnamefont {Zupancic}}, \bibinfo {author}
	  {\bibfnamefont {A.}~\bibnamefont {Baumg{\"a}rtner}}, \bibinfo {author}
	  {\bibfnamefont {A.}~\bibnamefont {Morales}}, \bibinfo {author} {\bibfnamefont
	  {W.}~\bibnamefont {Zheng}}, \bibinfo {author} {\bibfnamefont {N.~R.}\
	  \bibnamefont {Cooper}}, \bibinfo {author} {\bibfnamefont {T.}~\bibnamefont
	  {Donner}}, \ and\ \bibinfo {author} {\bibfnamefont {T.}~\bibnamefont
	  {Esslinger}},\ }\href {\doibase 10.1103/PhysRevResearch.3.L012024} {\bibfield
	   {journal} {\bibinfo  {journal} {Phys. Rev. Research}\ }\textbf {\bibinfo
	  {volume} {3}},\ \bibinfo {pages} {L012024} (\bibinfo {year}
	  {2021})}\BibitemShut {NoStop}%
	\bibitem [{\citenamefont {Chen}\ \emph {et~al.}(2014)\citenamefont {Chen},
	  \citenamefont {Yu},\ and\ \citenamefont {Zhai}}]{ThyFmSr@CY.2014}%
	  \BibitemOpen
	  \bibfield  {author} {\bibinfo {author} {\bibfnamefont {Y.}~\bibnamefont
	  {Chen}}, \bibinfo {author} {\bibfnamefont {Z.}~\bibnamefont {Yu}}, \ and\
	  \bibinfo {author} {\bibfnamefont {H.}~\bibnamefont {Zhai}},\ }\href {\doibase
	  10.1103/PhysRevLett.112.143004} {\bibfield  {journal} {\bibinfo  {journal}
	  {Phys. Rev. Lett.}\ }\textbf {\bibinfo {volume} {112}},\ \bibinfo {pages}
	  {143004} (\bibinfo {year} {2014})}\BibitemShut {NoStop}%
	\bibitem [{\citenamefont {Keeling}\ \emph {et~al.}(2014)\citenamefont
	  {Keeling}, \citenamefont {Bhaseen},\ and\ \citenamefont
	  {Simons}}]{ThyFmSr@Keeling.2014}%
	  \BibitemOpen
	  \bibfield  {author} {\bibinfo {author} {\bibfnamefont {J.}~\bibnamefont
	  {Keeling}}, \bibinfo {author} {\bibfnamefont {M.~J.}\ \bibnamefont
	  {Bhaseen}}, \ and\ \bibinfo {author} {\bibfnamefont {B.~D.}\ \bibnamefont
	  {Simons}},\ }\href {\doibase 10.1103/PhysRevLett.112.143002} {\bibfield
	  {journal} {\bibinfo  {journal} {Phys. Rev. Lett.}\ }\textbf {\bibinfo
	  {volume} {112}},\ \bibinfo {pages} {143002} (\bibinfo {year}
	  {2014})}\BibitemShut {NoStop}%
	\bibitem [{\citenamefont {Piazza}\ and\ \citenamefont
	  {Strack}(2014)}]{ThyFmSr@Piazza.2014}%
	  \BibitemOpen
	  \bibfield  {author} {\bibinfo {author} {\bibfnamefont {F.}~\bibnamefont
	  {Piazza}}\ and\ \bibinfo {author} {\bibfnamefont {P.}~\bibnamefont
	  {Strack}},\ }\href {\doibase 10.1103/PhysRevLett.112.143003} {\bibfield
	  {journal} {\bibinfo  {journal} {Phys. Rev. Lett.}\ }\textbf {\bibinfo
	  {volume} {112}},\ \bibinfo {pages} {143003} (\bibinfo {year}
	  {2014})}\BibitemShut {NoStop}%
	\bibitem [{\citenamefont {Chen}\ \emph {et~al.}(2015)\citenamefont {Chen},
	  \citenamefont {Zhai},\ and\ \citenamefont {Yu}}]{ThyFmSr@CY.2015}%
	  \BibitemOpen
	  \bibfield  {author} {\bibinfo {author} {\bibfnamefont {Y.}~\bibnamefont
	  {Chen}}, \bibinfo {author} {\bibfnamefont {H.}~\bibnamefont {Zhai}}, \ and\
	  \bibinfo {author} {\bibfnamefont {Z.}~\bibnamefont {Yu}},\ }\href {\doibase
	  10.1103/PhysRevA.91.021602} {\bibfield  {journal} {\bibinfo  {journal} {Phys.
	  Rev. A}\ }\textbf {\bibinfo {volume} {91}},\ \bibinfo {pages} {021602}
	  (\bibinfo {year} {2015})}\BibitemShut {NoStop}%
	\bibitem [{\citenamefont {Kollath}\ \emph {et~al.}(2016)\citenamefont
	  {Kollath}, \citenamefont {Sheikhan}, \citenamefont {Wolff},\ and\
	  \citenamefont {Brennecke}}]{ThyFmSr@Brennecke.2016}%
	  \BibitemOpen
	  \bibfield  {author} {\bibinfo {author} {\bibfnamefont {C.}~\bibnamefont
	  {Kollath}}, \bibinfo {author} {\bibfnamefont {A.}~\bibnamefont {Sheikhan}},
	  \bibinfo {author} {\bibfnamefont {S.}~\bibnamefont {Wolff}}, \ and\ \bibinfo
	  {author} {\bibfnamefont {F.}~\bibnamefont {Brennecke}},\ }\href {\doibase
	  10.1103/PhysRevLett.116.060401} {\bibfield  {journal} {\bibinfo  {journal}
	  {Phys. Rev. Lett.}\ }\textbf {\bibinfo {volume} {116}},\ \bibinfo {pages}
	  {060401} (\bibinfo {year} {2016})}\BibitemShut {NoStop}%
	\bibitem [{\citenamefont {Sheikhan}\ \emph {et~al.}(2016)\citenamefont
	  {Sheikhan}, \citenamefont {Brennecke},\ and\ \citenamefont
	  {Kollath}}]{ThyFmSr@Kollath.2016}%
	  \BibitemOpen
	  \bibfield  {author} {\bibinfo {author} {\bibfnamefont {A.}~\bibnamefont
	  {Sheikhan}}, \bibinfo {author} {\bibfnamefont {F.}~\bibnamefont {Brennecke}},
	  \ and\ \bibinfo {author} {\bibfnamefont {C.}~\bibnamefont {Kollath}},\ }\href
	  {\doibase 10.1103/PhysRevA.94.061603} {\bibfield  {journal} {\bibinfo
	  {journal} {Phys. Rev. A}\ }\textbf {\bibinfo {volume} {94}},\ \bibinfo
	  {pages} {061603} (\bibinfo {year} {2016})}\BibitemShut {NoStop}%
	\bibitem [{\citenamefont {Mivehvar}\ \emph {et~al.}(2017)\citenamefont
	  {Mivehvar}, \citenamefont {Ritsch},\ and\ \citenamefont
	  {Piazza}}]{ThyFmSr@Piazza.2017}%
	  \BibitemOpen
	  \bibfield  {author} {\bibinfo {author} {\bibfnamefont {F.}~\bibnamefont
	  {Mivehvar}}, \bibinfo {author} {\bibfnamefont {H.}~\bibnamefont {Ritsch}}, \
	  and\ \bibinfo {author} {\bibfnamefont {F.}~\bibnamefont {Piazza}},\ }\href
	  {\doibase 10.1103/PhysRevLett.118.073602} {\bibfield  {journal} {\bibinfo
	  {journal} {Phys. Rev. Lett.}\ }\textbf {\bibinfo {volume} {118}},\ \bibinfo
	  {pages} {073602} (\bibinfo {year} {2017})}\BibitemShut {NoStop}%
	\bibitem [{\citenamefont {Yu}\ \emph {et~al.}(2018)\citenamefont {Yu},
	  \citenamefont {Pan}, \citenamefont {Liu}, \citenamefont {Zhang},\ and\
	  \citenamefont {Yi}}]{ThyFmSr@YW.2018}%
	  \BibitemOpen
	  \bibfield  {author} {\bibinfo {author} {\bibfnamefont {D.}~\bibnamefont
	  {Yu}}, \bibinfo {author} {\bibfnamefont {J.-S.}\ \bibnamefont {Pan}},
	  \bibinfo {author} {\bibfnamefont {X.-J.}\ \bibnamefont {Liu}}, \bibinfo
	  {author} {\bibfnamefont {W.}~\bibnamefont {Zhang}}, \ and\ \bibinfo {author}
	  {\bibfnamefont {W.}~\bibnamefont {Yi}},\ }\href {\doibase
	  10.1007/s11467-017-0695-5} {\bibfield  {journal} {\bibinfo  {journal} {Front.
	  Phys.}\ }\textbf {\bibinfo {volume} {13}},\ \bibinfo {pages} {136701}
	  (\bibinfo {year} {2018})}\BibitemShut {NoStop}%
	\bibitem [{\citenamefont {Colella}\ \emph {et~al.}(2019)\citenamefont
	  {Colella}, \citenamefont {Ostermann}, \citenamefont {Niedenzu}, \citenamefont
	  {Mivehvar},\ and\ \citenamefont {Ritsch}}]{ThyFmSr@Ritsch.2019}%
	  \BibitemOpen
	  \bibfield  {author} {\bibinfo {author} {\bibfnamefont {E.}~\bibnamefont
	  {Colella}}, \bibinfo {author} {\bibfnamefont {S.}~\bibnamefont {Ostermann}},
	  \bibinfo {author} {\bibfnamefont {W.}~\bibnamefont {Niedenzu}}, \bibinfo
	  {author} {\bibfnamefont {F.}~\bibnamefont {Mivehvar}}, \ and\ \bibinfo
	  {author} {\bibfnamefont {H.}~\bibnamefont {Ritsch}},\ }\href {\doibase
	  10.1088/1367-2630/ab151e} {\bibfield  {journal} {\bibinfo  {journal} {New J.
	  Phys.}\ }\textbf {\bibinfo {volume} {21}},\ \bibinfo {pages} {043019}
	  (\bibinfo {year} {2019})}\BibitemShut {NoStop}%
	\bibitem [{\citenamefont {Zhang}\ \emph {et~al.}(2021)\citenamefont {Zhang},
	  \citenamefont {Chen}, \citenamefont {Wu}, \citenamefont {Wang}, \citenamefont
	  {Fan}, \citenamefont {Deng},\ and\ \citenamefont {Wu}}]{ExpFmSr@WHB.2021}%
	  \BibitemOpen
	  \bibfield  {author} {\bibinfo {author} {\bibfnamefont {X.}~\bibnamefont
	  {Zhang}}, \bibinfo {author} {\bibfnamefont {Y.}~\bibnamefont {Chen}},
	  \bibinfo {author} {\bibfnamefont {Z.}~\bibnamefont {Wu}}, \bibinfo {author}
	  {\bibfnamefont {J.}~\bibnamefont {Wang}}, \bibinfo {author} {\bibfnamefont
	  {J.}~\bibnamefont {Fan}}, \bibinfo {author} {\bibfnamefont {S.}~\bibnamefont
	  {Deng}}, \ and\ \bibinfo {author} {\bibfnamefont {H.}~\bibnamefont {Wu}},\
	  }\href {\doibase 10.1126/science.abd4385} {\bibfield  {journal} {\bibinfo
	  {journal} {Science}\ }\textbf {\bibinfo {volume} {373}},\ \bibinfo {pages}
	  {1359} (\bibinfo {year} {2021})}\BibitemShut {NoStop}%
	\bibitem [{\citenamefont {Bhaseen}\ \emph {et~al.}(2012)\citenamefont
	  {Bhaseen}, \citenamefont {Mayoh}, \citenamefont {Simons},\ and\ \citenamefont
	  {Keeling}}]{ThyBsUns@Keeling.2012}%
	  \BibitemOpen
	  \bibfield  {author} {\bibinfo {author} {\bibfnamefont {M.~J.}\ \bibnamefont
	  {Bhaseen}}, \bibinfo {author} {\bibfnamefont {J.}~\bibnamefont {Mayoh}},
	  \bibinfo {author} {\bibfnamefont {B.~D.}\ \bibnamefont {Simons}}, \ and\
	  \bibinfo {author} {\bibfnamefont {J.}~\bibnamefont {Keeling}},\ }\href
	  {\doibase 10.1103/PhysRevA.85.013817} {\bibfield  {journal} {\bibinfo
	  {journal} {Phys. Rev. A}\ }\textbf {\bibinfo {volume} {85}},\ \bibinfo
	  {pages} {013817} (\bibinfo {year} {2012})}\BibitemShut {NoStop}%
	\bibitem [{\citenamefont {Piazza}\ and\ \citenamefont
	  {Ritsch}(2015)}]{OTC@Piazza.2015}%
	  \BibitemOpen
	  \bibfield  {author} {\bibinfo {author} {\bibfnamefont {F.}~\bibnamefont
	  {Piazza}}\ and\ \bibinfo {author} {\bibfnamefont {H.}~\bibnamefont
	  {Ritsch}},\ }\href {\doibase 10.1103/PhysRevLett.115.163601} {\bibfield
	  {journal} {\bibinfo  {journal} {Phys. Rev. Lett.}\ }\textbf {\bibinfo
	  {volume} {115}},\ \bibinfo {pages} {163601} (\bibinfo {year}
	  {2015})}\BibitemShut {NoStop}%
	\bibitem [{\citenamefont {Zheng}\ and\ \citenamefont {Cooper}(2016)}]{ZW@2016}%
	  \BibitemOpen
	  \bibfield  {author} {\bibinfo {author} {\bibfnamefont {W.}~\bibnamefont
	  {Zheng}}\ and\ \bibinfo {author} {\bibfnamefont {N.~R.}\ \bibnamefont
	  {Cooper}},\ }\href {\doibase 10.1103/PhysRevLett.117.175302} {\bibfield
	  {journal} {\bibinfo  {journal} {Phys. Rev. Lett.}\ }\textbf {\bibinfo
	  {volume} {117}},\ \bibinfo {pages} {175302} (\bibinfo {year}
	  {2016})}\BibitemShut {NoStop}%
	\bibitem [{\citenamefont {Kirton}\ and\ \citenamefont
	  {Keeling}(2018)}]{DickeCTC@Keeling.2018}%
	  \BibitemOpen
	  \bibfield  {author} {\bibinfo {author} {\bibfnamefont {P.}~\bibnamefont
	  {Kirton}}\ and\ \bibinfo {author} {\bibfnamefont {J.}~\bibnamefont
	  {Keeling}},\ }\href {\doibase 10.1088/1367-2630/aaa11d} {\bibfield  {journal}
	  {\bibinfo  {journal} {New J. Phys.}\ }\textbf {\bibinfo {volume} {20}},\
	  \bibinfo {pages} {015009} (\bibinfo {year} {2018})}\BibitemShut {NoStop}%
	\bibitem [{\citenamefont {Ke{\ss}ler}\ \emph {et~al.}(2019)\citenamefont
	  {Ke{\ss}ler}, \citenamefont {Cosme}, \citenamefont {Hemmerling},
	  \citenamefont {Mathey},\ and\ \citenamefont
	  {Hemmerich}}]{CTC@Hemmerich.2019}%
	  \BibitemOpen
	  \bibfield  {author} {\bibinfo {author} {\bibfnamefont {H.}~\bibnamefont
	  {Ke{\ss}ler}}, \bibinfo {author} {\bibfnamefont {J.~G.}\ \bibnamefont
	  {Cosme}}, \bibinfo {author} {\bibfnamefont {M.}~\bibnamefont {Hemmerling}},
	  \bibinfo {author} {\bibfnamefont {L.}~\bibnamefont {Mathey}}, \ and\ \bibinfo
	  {author} {\bibfnamefont {A.}~\bibnamefont {Hemmerich}},\ }\href {\doibase
	  10.1103/PhysRevA.99.053605} {\bibfield  {journal} {\bibinfo  {journal} {Phys.
	  Rev. A}\ }\textbf {\bibinfo {volume} {99}},\ \bibinfo {pages} {053605}
	  (\bibinfo {year} {2019})}\BibitemShut {NoStop}%
	\bibitem [{\citenamefont {Chiacchio}\ and\ \citenamefont
	  {Nunnenkamp}(2019)}]{ThyBsUnst@Nunnenkamp.2019}%
	  \BibitemOpen
	  \bibfield  {author} {\bibinfo {author} {\bibfnamefont {E.~I.~R.}\
	  \bibnamefont {Chiacchio}}\ and\ \bibinfo {author} {\bibfnamefont
	  {A.}~\bibnamefont {Nunnenkamp}},\ }\href {\doibase
	  10.1103/PhysRevLett.122.193605} {\bibfield  {journal} {\bibinfo  {journal}
	  {Phys. Rev. Lett.}\ }\textbf {\bibinfo {volume} {122}},\ \bibinfo {pages}
	  {193605} (\bibinfo {year} {2019})}\BibitemShut {NoStop}%
	\bibitem [{\citenamefont {Bu{\v c}a}\ and\ \citenamefont
	  {Jaksch}(2019)}]{OTC2@Jaksch.2019}%
	  \BibitemOpen
	  \bibfield  {author} {\bibinfo {author} {\bibfnamefont {B.}~\bibnamefont
	  {Bu{\v c}a}}\ and\ \bibinfo {author} {\bibfnamefont {D.}~\bibnamefont
	  {Jaksch}},\ }\href {\doibase 10.1103/PhysRevLett.123.260401} {\bibfield
	  {journal} {\bibinfo  {journal} {Phys. Rev. Lett.}\ }\textbf {\bibinfo
	  {volume} {123}},\ \bibinfo {pages} {260401} (\bibinfo {year}
	  {2019})}\BibitemShut {NoStop}%
	\bibitem [{\citenamefont {Tuquero}\ \emph {et~al.}(2022)\citenamefont
	  {Tuquero}, \citenamefont {Skulte}, \citenamefont {Mathey},\ and\
	  \citenamefont {Cosme}}]{OTC@Tuquero.2022}%
	  \BibitemOpen
	  \bibfield  {author} {\bibinfo {author} {\bibfnamefont {R.~J.~L.}\
	  \bibnamefont {Tuquero}}, \bibinfo {author} {\bibfnamefont {J.}~\bibnamefont
	  {Skulte}}, \bibinfo {author} {\bibfnamefont {L.}~\bibnamefont {Mathey}}, \
	  and\ \bibinfo {author} {\bibfnamefont {J.~G.}\ \bibnamefont {Cosme}},\ }\href
	  {\doibase 10.1103/PhysRevA.105.043311} {\bibfield  {journal} {\bibinfo
	  {journal} {Phys. Rev. A}\ }\textbf {\bibinfo {volume} {105}},\ \bibinfo
	  {pages} {043311} (\bibinfo {year} {2022})}\BibitemShut {NoStop}%
	\bibitem [{\citenamefont {Zhang}\ \emph {et~al.}(2022)\citenamefont {Zhang},
	  \citenamefont {Dreon}, \citenamefont {Esslinger}, \citenamefont {Jaksch},
	  \citenamefont {Buca},\ and\ \citenamefont
	  {Donner}}]{ThyBsUns@Esslinger.2022}%
	  \BibitemOpen
	  \bibfield  {author} {\bibinfo {author} {\bibfnamefont {Z.}~\bibnamefont
	  {Zhang}}, \bibinfo {author} {\bibfnamefont {D.}~\bibnamefont {Dreon}},
	  \bibinfo {author} {\bibfnamefont {T.}~\bibnamefont {Esslinger}}, \bibinfo
	  {author} {\bibfnamefont {D.}~\bibnamefont {Jaksch}}, \bibinfo {author}
	  {\bibfnamefont {B.}~\bibnamefont {Buca}}, \ and\ \bibinfo {author}
	  {\bibfnamefont {T.}~\bibnamefont {Donner}},\ }\href@noop {} {\enquote
	  {\bibinfo {title} {Tunable non-equilibrium phase transitions between spatial
	  and temporal order through dissipation},}\ } (\bibinfo {year} {2022}),\
	  \Eprint {http://arxiv.org/abs/2205.01461} {arxiv:2205.01461} \BibitemShut
	  {NoStop}%
	\bibitem [{\citenamefont {Nie}\ and\ \citenamefont {Zheng}(2023)}]{Zheng.2023}%
	  \BibitemOpen
	  \bibfield  {author} {\bibinfo {author} {\bibfnamefont {X.}~\bibnamefont
	  {Nie}}\ and\ \bibinfo {author} {\bibfnamefont {W.}~\bibnamefont {Zheng}},\
	  }\href {\doibase 10.1103/PhysRevA.107.033311} {\bibfield  {journal} {\bibinfo
	   {journal} {Phys. Rev. A}\ }\textbf {\bibinfo {volume} {107}},\ \bibinfo
	  {pages} {033311} (\bibinfo {year} {2023})}\BibitemShut {NoStop}%
	\bibitem [{\citenamefont {Zupancic}\ \emph {et~al.}(2019)\citenamefont
	  {Zupancic}, \citenamefont {Dreon}, \citenamefont {Li}, \citenamefont
	  {Baumg{\"a}rtner}, \citenamefont {Morales}, \citenamefont {Zheng},
	  \citenamefont {Cooper}, \citenamefont {Esslinger},\ and\ \citenamefont
	  {Donner}}]{ExpBsUns@Esslinger.2019}%
	  \BibitemOpen
	  \bibfield  {author} {\bibinfo {author} {\bibfnamefont {P.}~\bibnamefont
	  {Zupancic}}, \bibinfo {author} {\bibfnamefont {D.}~\bibnamefont {Dreon}},
	  \bibinfo {author} {\bibfnamefont {X.}~\bibnamefont {Li}}, \bibinfo {author}
	  {\bibfnamefont {A.}~\bibnamefont {Baumg{\"a}rtner}}, \bibinfo {author}
	  {\bibfnamefont {A.}~\bibnamefont {Morales}}, \bibinfo {author} {\bibfnamefont
	  {W.}~\bibnamefont {Zheng}}, \bibinfo {author} {\bibfnamefont {N.~R.}\
	  \bibnamefont {Cooper}}, \bibinfo {author} {\bibfnamefont {T.}~\bibnamefont
	  {Esslinger}}, \ and\ \bibinfo {author} {\bibfnamefont {T.}~\bibnamefont
	  {Donner}},\ }\href {\doibase 10.1103/PhysRevLett.123.233601} {\bibfield
	  {journal} {\bibinfo  {journal} {Phys. Rev. Lett.}\ }\textbf {\bibinfo
	  {volume} {123}},\ \bibinfo {pages} {233601} (\bibinfo {year}
	  {2019})}\BibitemShut {NoStop}%
	\bibitem [{\citenamefont {Dogra}\ \emph {et~al.}(2019)\citenamefont {Dogra},
	  \citenamefont {Landini}, \citenamefont {Kroeger}, \citenamefont {Hruby},
	  \citenamefont {Donner},\ and\ \citenamefont
	  {Esslinger}}]{ExpBsUns2@Esslinger.2019}%
	  \BibitemOpen
	  \bibfield  {author} {\bibinfo {author} {\bibfnamefont {N.}~\bibnamefont
	  {Dogra}}, \bibinfo {author} {\bibfnamefont {M.}~\bibnamefont {Landini}},
	  \bibinfo {author} {\bibfnamefont {K.}~\bibnamefont {Kroeger}}, \bibinfo
	  {author} {\bibfnamefont {L.}~\bibnamefont {Hruby}}, \bibinfo {author}
	  {\bibfnamefont {T.}~\bibnamefont {Donner}}, \ and\ \bibinfo {author}
	  {\bibfnamefont {T.}~\bibnamefont {Esslinger}},\ }\href {\doibase
	  10.1126/science.aaw4465} {\bibfield  {journal} {\bibinfo  {journal}
	  {Science}\ }\textbf {\bibinfo {volume} {366}},\ \bibinfo {pages} {1496}
	  (\bibinfo {year} {2019})}\BibitemShut {NoStop}%
	\bibitem [{\citenamefont {Kongkhambut}\ \emph {et~al.}(2022)\citenamefont
	  {Kongkhambut}, \citenamefont {Skulte}, \citenamefont {Mathey}, \citenamefont
	  {Cosme}, \citenamefont {Hemmerich},\ and\ \citenamefont
	  {Ke{\ss}ler}}]{CTC@Hemmerich.2022}%
	  \BibitemOpen
	  \bibfield  {author} {\bibinfo {author} {\bibfnamefont {P.}~\bibnamefont
	  {Kongkhambut}}, \bibinfo {author} {\bibfnamefont {J.}~\bibnamefont {Skulte}},
	  \bibinfo {author} {\bibfnamefont {L.}~\bibnamefont {Mathey}}, \bibinfo
	  {author} {\bibfnamefont {J.~G.}\ \bibnamefont {Cosme}}, \bibinfo {author}
	  {\bibfnamefont {A.}~\bibnamefont {Hemmerich}}, \ and\ \bibinfo {author}
	  {\bibfnamefont {H.}~\bibnamefont {Ke{\ss}ler}},\ }\href {\doibase
	  10.1126/science.abo3382} {\bibfield  {journal} {\bibinfo  {journal}
	  {Science}\ }\textbf {\bibinfo {volume} {377}},\ \bibinfo {pages} {670}
	  (\bibinfo {year} {2022})}\BibitemShut {NoStop}%
	\bibitem [{\citenamefont {Dreon}\ \emph {et~al.}(2022)\citenamefont {Dreon},
	  \citenamefont {Baumg{\"a}rtner}, \citenamefont {Li}, \citenamefont
	  {Hertlein}, \citenamefont {Esslinger},\ and\ \citenamefont
	  {Donner}}]{ExpBsUns@Esslinger.2022}%
	  \BibitemOpen
	  \bibfield  {author} {\bibinfo {author} {\bibfnamefont {D.}~\bibnamefont
	  {Dreon}}, \bibinfo {author} {\bibfnamefont {A.}~\bibnamefont
	  {Baumg{\"a}rtner}}, \bibinfo {author} {\bibfnamefont {X.}~\bibnamefont {Li}},
	  \bibinfo {author} {\bibfnamefont {S.}~\bibnamefont {Hertlein}}, \bibinfo
	  {author} {\bibfnamefont {T.}~\bibnamefont {Esslinger}}, \ and\ \bibinfo
	  {author} {\bibfnamefont {T.}~\bibnamefont {Donner}},\ }\href {\doibase
	  10.1038/s41586-022-04970-0} {\bibfield  {journal} {\bibinfo  {journal}
	  {Nature}\ }\textbf {\bibinfo {volume} {608}},\ \bibinfo {pages} {494}
	  (\bibinfo {year} {2022})}\BibitemShut {NoStop}%
	\bibitem [{\citenamefont {Nataf}\ \emph {et~al.}(2012)\citenamefont {Nataf},
	  \citenamefont {Baksic},\ and\ \citenamefont {Ciuti}}]{Ciuti.2012}%
	  \BibitemOpen
	  \bibfield  {author} {\bibinfo {author} {\bibfnamefont {P.}~\bibnamefont
	  {Nataf}}, \bibinfo {author} {\bibfnamefont {A.}~\bibnamefont {Baksic}}, \
	  and\ \bibinfo {author} {\bibfnamefont {C.}~\bibnamefont {Ciuti}},\ }\href
	  {\doibase 10.1103/PhysRevA.86.013832} {\bibfield  {journal} {\bibinfo
	  {journal} {Phys. Rev. A}\ }\textbf {\bibinfo {volume} {86}},\ \bibinfo
	  {pages} {013832} (\bibinfo {year} {2012})}\BibitemShut {NoStop}%
	\bibitem [{\citenamefont {Vanderbilt}\ and\ \citenamefont
	  {{King-Smith}}(1993)}]{King-Smith.1993}%
	  \BibitemOpen
	  \bibfield  {author} {\bibinfo {author} {\bibfnamefont {D.}~\bibnamefont
	  {Vanderbilt}}\ and\ \bibinfo {author} {\bibfnamefont {R.~D.}\ \bibnamefont
	  {{King-Smith}}},\ }\href {\doibase 10.1103/PhysRevB.48.4442} {\bibfield
	  {journal} {\bibinfo  {journal} {Phys. Rev. B}\ }\textbf {\bibinfo {volume}
	  {48}},\ \bibinfo {pages} {4442} (\bibinfo {year} {1993})}\BibitemShut
	  {NoStop}%
	\bibitem [{\citenamefont {Resta}(1994)}]{Resta.1994}%
	  \BibitemOpen
	  \bibfield  {author} {\bibinfo {author} {\bibfnamefont {R.}~\bibnamefont
	  {Resta}},\ }\href {\doibase 10.1103/RevModPhys.66.899} {\bibfield  {journal}
	  {\bibinfo  {journal} {Rev. Mod. Phys.}\ }\textbf {\bibinfo {volume} {66}},\
	  \bibinfo {pages} {899} (\bibinfo {year} {1994})}\BibitemShut {NoStop}%
	\bibitem [{\citenamefont {Xiao}\ \emph {et~al.}(2010)\citenamefont {Xiao},
	  \citenamefont {Chang},\ and\ \citenamefont {Niu}}]{Niu.2010}%
	  \BibitemOpen
	  \bibfield  {author} {\bibinfo {author} {\bibfnamefont {D.}~\bibnamefont
	  {Xiao}}, \bibinfo {author} {\bibfnamefont {M.-C.}\ \bibnamefont {Chang}}, \
	  and\ \bibinfo {author} {\bibfnamefont {Q.}~\bibnamefont {Niu}},\ }\href
	  {\doibase 10.1103/RevModPhys.82.1959} {\bibfield  {journal} {\bibinfo
	  {journal} {Rev. Mod. Phys.}\ }\textbf {\bibinfo {volume} {82}},\ \bibinfo
	  {pages} {1959} (\bibinfo {year} {2010})}\BibitemShut {NoStop}%
	\bibitem [{\citenamefont {Thouless}(1983)}]{Thouless.1983}%
	  \BibitemOpen
	  \bibfield  {author} {\bibinfo {author} {\bibfnamefont {D.~J.}\ \bibnamefont
	  {Thouless}},\ }\href {\doibase 10.1103/PhysRevB.27.6083} {\bibfield
	  {journal} {\bibinfo  {journal} {Phys. Rev. B}\ }\textbf {\bibinfo {volume}
	  {27}},\ \bibinfo {pages} {6083} (\bibinfo {year} {1983})}\BibitemShut
	  {NoStop}%
	\bibitem [{\citenamefont {Citro}\ and\ \citenamefont
	  {Aidelsburger}(2023)}]{Aidelsburger.2023}%
	  \BibitemOpen
	  \bibfield  {author} {\bibinfo {author} {\bibfnamefont {R.}~\bibnamefont
	  {Citro}}\ and\ \bibinfo {author} {\bibfnamefont {M.}~\bibnamefont
	  {Aidelsburger}},\ }\href {\doibase 10.1038/s42254-022-00545-0} {\bibfield
	  {journal} {\bibinfo  {journal} {Nat Rev Phys}\ }\textbf {\bibinfo {volume}
	  {5}},\ \bibinfo {pages} {87} (\bibinfo {year} {2023})}\BibitemShut {NoStop}%
	\bibitem [{\citenamefont {Nakajima}\ \emph {et~al.}(2016)\citenamefont
	  {Nakajima}, \citenamefont {Tomita}, \citenamefont {Taie}, \citenamefont
	  {Ichinose}, \citenamefont {Ozawa}, \citenamefont {Wang}, \citenamefont
	  {Troyer},\ and\ \citenamefont {Takahashi}}]{Takahashi.2016}%
	  \BibitemOpen
	  \bibfield  {author} {\bibinfo {author} {\bibfnamefont {S.}~\bibnamefont
	  {Nakajima}}, \bibinfo {author} {\bibfnamefont {T.}~\bibnamefont {Tomita}},
	  \bibinfo {author} {\bibfnamefont {S.}~\bibnamefont {Taie}}, \bibinfo {author}
	  {\bibfnamefont {T.}~\bibnamefont {Ichinose}}, \bibinfo {author}
	  {\bibfnamefont {H.}~\bibnamefont {Ozawa}}, \bibinfo {author} {\bibfnamefont
	  {L.}~\bibnamefont {Wang}}, \bibinfo {author} {\bibfnamefont {M.}~\bibnamefont
	  {Troyer}}, \ and\ \bibinfo {author} {\bibfnamefont {Y.}~\bibnamefont
	  {Takahashi}},\ }\href {\doibase 10.1038/nphys3622} {\bibfield  {journal}
	  {\bibinfo  {journal} {Nature Phys}\ }\textbf {\bibinfo {volume} {12}},\
	  \bibinfo {pages} {296} (\bibinfo {year} {2016})}\BibitemShut {NoStop}%
	\end{thebibliography}
%
	
\end{document}